\documentclass[a4paper,10pt,twocolumn]{article}
\usepackage[utf8]{inputenc}

\usepackage{bm}
\usepackage{graphicx}
\usepackage{amsfonts}
\usepackage{amsmath}
\usepackage{amssymb}
\usepackage{amsthm}
\usepackage{bbm}
\usepackage{setspace}
\usepackage{helvet}

\usepackage{subfig}
\usepackage{color}
\definecolor{Red}{rgb}{0.9,0.0,0.1}

\makeatletter
\def\imod#1{\allowbreak\mkern10mu({\operator@font mod}\,\,#1)}
\makeatother

\title{Quantum Monte Carlo Study of Long-Range Transverse-Field Ising Models on the Triangular Lattice}
\author{Stephan Humeniuk \\ Institute for Theoretical Physics III, University of Stuttgart, Germany}

\begin{document}

\onecolumn
\maketitle

\begin{abstract}
Motivated by recent experiments with a Penning ion trap quantum simulator, we perform numerically exact 
Stochastic Series Expansion quantum Monte Carlo simulations of long-range transverse-field Ising models on a triangular lattice for different decay powers $\alpha$
of the interactions. The phase boundary for the ferromagnet is obtained as a function of $\alpha$. 
For antiferromagnetic interactions, there is strong indication that the transverse field stabilizes a clock ordered phase with sublattice magnetization 
$(M,-\frac{M}{2}, -\frac{M}{2})$ with unsaturated $M < 1$ in a process known as ``order by disorder'' similar to the nearest neighbour antiferromagnet 
on the triangular lattice. Connecting the known limiting cases of nearest neighbour
and infinite-range interactions, a semiquantitative phase diagram is obtained.
Magnetization curves for the ferromagnet for experimentally relevant system sizes and with open boundary conditions are presented.
\end{abstract}

\twocolumn
\section{Introduction}

The large ground state degeneracy which is the hallmark of geometrically frustrated magnets \cite{Lacroix2010, Diep2004} can give rise to emergent phenomena
governed by degrees of freedom that are quite distinct from those of the underlying 
spin system. An exotic phase that is particularly sought after in frustrated magnets 
is the famous spin liquid state which was first proposed as a possible ground state for the triangular lattice Heisenberg
model \cite{Anderson1973,Fazekas1974} and has been found e.g. in a charge-transfer salt, adjacent
to a superconducting phase \cite{Kurosaki2005}. The study of frustrated magnetism is plagued in many cases by the sign problem
which makes quantum Monte Carlo (QMC) simulations infeasible, while an exact diagonalization of the Hamiltonian 
for sufficiently large systems is impossible due to the exponential increase
of the Hilbert space with system size. Therefore the idea of a quantum simulator has been put forward \cite{Feynman1982, LLoyd1996}, i.e.
using a well-controlled physical system to emulate another physical system, adding complexity step by step.

In reference \cite{Britton2012} a Penning trap quantum simulator is described which is expected to simulate a transverse field 
Ising model with Hamiltonian 
\begin{equation}
 \hat{H} = \frac{1}{2}\sum_{ij} J_{ij} \hat{S}_i^z \hat{S}_j^z  - \Gamma \sum_i \hat{S}_i^x 
 \label{eq:H_TIF}
\end{equation} 
with tunable long-range interactions \mbox{$J_{ij} = J | {\bf r}_i - {\bf r}_j |^{-\alpha}$}.
We summarize the setup in the following:
When a collection of ions is laser-cooled in a Penning trap, below a certain temperature the ions undergo
a structural phase transition from a plasma to a Wigner crystal that is stabilized by their mutual Coulomb repulsion.
For a strong trapping potential in z-direction the system becomes two-dimensional and the corresponding
Wigner crystal is a triangular lattice. The number of trapped ions in a Penning trap ranges from a few to a few millions;
in \cite{Britton2012} the creation of a triangular Wigner crystal with $\approx$ 300 $^{9}\text{Be}^+$ ions was reported. 
There, the valence electron of each ion serves as a qubit, which is represented by the spin-$\frac{1}{2}$ operators $(\hat{S}^x_i, \hat{S}^y_i, \hat{S}^z_i)$
in Eq. \eqref{eq:H_TIF}. 
It can be driven by applying microwave radiation,
which corresponds to an effective magnetic field that can be tuned to have both transverse and longitudinal components.
The use of a spin-dependent optical dipole force allows to engineer a long-range Ising-type
spin-spin interaction $J_{i,j} \propto |r_i - r_j|^{-\alpha}$ between the internal two-level systems of the ions \cite{Porras2004}, where the exponent $\alpha$ 
can be continuously varied between 0 and 3 (dipolar decay) by adjusting a laser detuning parameter. The mechanism by which this interaction arises is 
naturally long-range and antiferromagnetic (AFM) if the direction of the spin-dependent optical dipole-force is perpendicular to the $2d$ Coulomb crystal
so that ions with opposite spin move away from each other in the $z$-direction thereby minimizing their Coulomb repulsion.
Ferromagnetic (FM) interactions can be generated by additionally adjusting a detuning from the eigenfrequencies of normal modes of the Coulomb crystal. 
The goal of this paper is to map out the ground state phase diagram of this effective spin model as a function of $\alpha$ and $\Gamma$ 
for both ferromagnetic and antiferromagnetic interactions,
which may help to benchmark such a potential quantum simulator.

\emph{Long-range interactions:}
Long-range systems are interesting as they exhibit a number of thermodynamic and dynamical peculiarities \cite{Dyson1969, Dauxois2002, Mukamel2009, Worm2013, Peter2012}. 
% (such as diverging equilibration timescales and pre-thermalization plateux for $0 \le \alpha \le d/2$ \cite{Worm2013} and 
% ensemble inequivalence \cite{Kastner2010}) and they defy basic tenets of the statistical mechanics of short-range systems \cite{Dyson1969}.
While all ground states of gapped short-range Hamiltonians obey the ``area law'' for the entanglement entropy, a violation thereof was recently found in 
the long-range transverse-field Ising chain \cite{Koffel2012} along with algebraically decaying correlation functions in a gapped phase
 (see also \cite{Peter2012} for this last point).
Furthermore, apart from ion traps, one of the best realizations of the transverse-field Ising model, the Ising ferromagnet LiHoF$_4$ \cite{Chakraborty2004},
is actually long-range with a dipolar decay of the interactions.

Few works have dealt with the competition of long-range and AFM interactions in 2d quantum spin models \cite{Peter2012}.
On a bipartite lattice, long-range AFM interactions induce only weak frustration. The triangular lattice, on the other hand,
is already frustrated, giving rise, in the case of the short-range Ising AFM, to a disordered classical ground state manifold
with the long-range interactions leading - in principle - to additional frustration. 
The minimum energy configurations of the classical model \eqref{eq:H_TIF} with $\Gamma=0$ on the triangular lattice 
may depend sensitively on $\alpha$ 
and computing them may amount to a difficult optimization problem. To the best of our knowledge
the classical phase diagram of \eqref{eq:H_TIF} as a function of decay exponent 
$\alpha$ is not known.
% In the context of ultrathin magnetic films minimum energy configurations for dipolar AFM Ising interactions have been obtained
% for various lattice types in 2d (see for example \cite{MacIsaac1995}), however, always including a ferromagnetic nearest-neighbour exchange coupling as a 
% model for the formation of magnetic domains, which makes these systems very different from the one studied here.
It will be shown in the following that including quantum fluctuations through
a transverse field, however, stabilizes a phase that extends over a range of decay exponents $\alpha$, thus 
merging potentially different classical ground states. 

\emph{Order by disorder:}
Let us first summarize the known results for the short-range case, starting from the classical triangular Ising antiferromagnet
and moving on to ferromangetically stacked layers of antiferromagnets, which - by virtue of a quantum-to-classical
mapping - corresponds to a transverse field quantum Ising model, which is the main subject of our interest.
The phase diagram of the classical triangular Ising antiferromagnet, which is obtained by setting $\Gamma = 0$ in
equation \eqref{eq:H_TIF}, is well-known from the work of Wannier \cite{Wannier1950} and Stephenson \cite{Stephenson1970}: It is disordered at all temperatures with a macroscopic ground state degeneracy \cite{Wannier1950} and critical spin-spin correlations
at $T=0$ where $\langle S^z_0 S^z_r \rangle$ decreases asymtotically as $r^{-1/2}$ \cite{Stephenson1970}. 
The ground state degeneracy can be explained by looking at the Fourier transformation, $\tilde{J}({\bf q})$, of the Ising interactions on the triangular lattice and 
taking into account the nature of the Ising spins.
While 
\begin{equation}
 \tilde{J}({\bf q}) = J \left[\cos q_x + 2 \cos\left(\frac{q_x}{2}\right) \cos \left( \frac{\sqrt{3} q_y}{2} \right) \right]
\end{equation}
has minima of $-3J/2$ at the corners of the hexagonal Brillouin zone, at ${\bf Q}_{\pm} = (\pm \frac{4 \pi}{3} , 0)$ and equivalent points,
such ordering vectors are not compatible with the hard-spin constraint of classical Ising spins \cite{Alexander1980}. The lowest possible energy
per frustrated triangle can be realized in a multitude of ways so that a macroscopic ground state 
degeneracy arises. 
The situation changes for stacked triangular antiferromagnets \cite{Blankschtein1984}.
If several triangular layers are stacked ferromagnetically on top of each other, chains of spins in the stacking direction can be combined 
to form an averaged macrospin. Then, for finite temperature, the spins, which are coupled ferromagnetically within a chain, 
fluctuate and the averaging removes the hard-spin constraint
so that the 2d system of chains can settle into the minima at ordering vectors ${\bf Q}_{\pm}$. This is the classical, i.e. finite-temperature induced version of 
the phenomenon ``order by disorder''.

The transverse field quantum Ising model can be mapped by the Trotter-Suzuki formalism
to a ferromagnetically stacked classical Ising model so that the statements made above for stacked triangular magnets carry over to the $2d$
transverse field Ising model on the triangular lattice \cite{Isakov2003}. 
The mechanism that is responsible for the appearance of clock order in a transverse field is the quantum version of ``order by disorder'':
Quantum fluctuations induced by the transverse field stabilize those states from the classical ground state manifold
which can lower their energy the most by resonance processes. Thus, the exponential degeneracy of the ground state manifold is lifted 
and - since the favoured states tend to be regularly structured spin configurations - an ordered state emerges.
Which resonance processes are possible
within the ground state manifold (and to which order in $\Gamma$) is determined by the lattice structure and the interactions \cite{Moessner2001}.
For fully frustrated systems, the action of the transverse field to lowest order within the ground state manifold is equivalent to $S_i^{+} S_j^{-} + h.c.$
because one always needs to flip two spins in order to get back to the ground state manifold.

\emph{XY order parameter:}
The ordering vectors ${\bf Q}_{\pm} = (\pm 4\pi/3, 0)$ of the critical modes correspond to inequivalent points in the hexagonal Brillouin zone.
This implies a two-component order parameter.
A Landau-Ginzburg-Wilson analysis \cite{Alexander1980, Blankschtein1984} found an XY action with an XY symmetry breaking clock anisotropy term 
\begin{multline}
 H_{\text{LGW}} = \sum_{{\bf q}} (r + {\bf q}^2)m^2 + \\
 u_4 \sum_4 m^4 + u_6 \sum_6 m^6 + v_6 \sum_6 m^6 \cos(6 \vartheta)
 \label{eq:LGW_Hamiltonian}
\end{multline}

which,
depending on the sign of the coefficient $v_6$, selects between two different three-sublattice ordered states according
to \cite{Blankschtein1984} \mbox{$\langle s_j \rangle \propto M \cos ({\bf Q}_{+}\cdot {\bf r}_j + \theta)$}, namely a ferrimagnetic state $(1, -\frac{1}{2}, -\frac{1}{2})$ 
for $\theta = \vartheta = 0$ and 
a partially disordered antiferromagnetic state $(1, -1, 0)$ for $\theta = \vartheta = \pi/6$. Both states are six-fold degenerate,
which can be seen by relabeling the sublattices and by making use of spin-inversion symmetry. 
Monte Carlo simulations \cite{Blankschtein1984,Netz1991} showed that at intermediate temperatures 
the stacked triangular antiferromagnet orders according to $(1, -1, 0)$ and at low temperatures 
approximately according to $(1, -\frac{1}{2}, -\frac{1}{2})$ with possibly unsaturated magnetization.
However, the nature of the low-temperature phase has been the subject of controversy 
\cite{Coppersmith1985,Heinonen1989,Kim1990,Zukovic2013}.

The fact that the frustration on the triangular lattice generates
an XY order parameter has several consequences:
With the clock term in \eqref{eq:LGW_Hamiltonian} being dangerously irrelevant, the transition from the paramagnetic to the clock ordered phase
is believed to be in the 3d XY universality class \cite{Blankschtein1984, Isakov2003}. The location of this quantum 
critical point was determined to be at $\Gamma_c /J = 1.65 \pm 0.05$ \cite{Isakov2003}.
The Kosterlitz-Thouless transition from a clock-ordered phase with sublattice magnetization $(1, -1, 0)$ 
to a paramagnetic phase via an extended critical phase, 
which occurs as a consequence of a finite-temperature induced dimensional crossover in the $(d+1)$-dimensional quantum system, 
was also investigated in Ref. \cite{Isakov2003}. 
The possibility of a Kosterlitz-Thouless transition at 
finite temperature in the long-range system is beyond the scope of this work.

The structure of this paper is as follows: Section 2 gives a mean-field analysis of both the ferromagnetic and antiferromagnetic problem in a transverse field. 
In Section 3 we briefly discuss
the variant of the Stochastic Series Expansion QMC method that we have used. Section 4 contains
results on the long-range ferromagnet and antiferromagnet, respectively. 
The main results are summarized in Section 5.

\section{Mean field theory}
\label{sec:mean-field}
It is well-known that for the infinitely coordinated Ising model, that is with weak constant interactions $J_{ij}= \frac{J}{N}$, where $N$ is the number of sites,
the saddle point approximation, which maps the system to a mean-field problem, becomes exact. It is intuitively clear that for long-range systems mean-field
theory provides a good description. However, this is only true for FM interactions since long-range AFM interactions lead to additional frustration.
We start from the Hamiltonian

\begin{equation}
 \hat{H} = \frac{1}{2} \sum_{ i,j } J_{ij} \hat{S}_i^z \hat{S}_j^z - \Gamma \sum_i \hat{S}_i^x - h_{\parallel} \sum_i \hat{S}_i^z
\end{equation}
where for the sake of completeness we have added a longitudinal field $h_{\parallel}$.
Writing the spin operator in terms of small fluctuations around an average value, $ \hat{S}_i^z = \langle S_i^z \rangle + \delta \hat{S}_i^z \equiv m +  \delta \hat{S}_i^z$,
and neglecting second-order fluctuations, a mean-field Hamiltonian can be obtained
\begin{equation}
 \hat{H}_{MF} = \frac{1}{2} N \tilde{J}(0) m^2 - (\tilde{J}(0) m + h_{\parallel}) \sum_{i} \hat{S}_i^z - \Gamma \sum_{i} \hat{S}_i^x,
\end{equation} where $\tilde{J}({\bf q}) = \sum_{{\bf R}}  J({\bf R}) \exp(-i {\bf q} \cdot {\bf R})$ is the Fourier transform of the interactions.
Thus, dropping an overall constant, the thermodynamics reduces to that of paramagnetic spins in a longitudinal and transverse field. 
The single-spin Hamiltonian can easily be diagonalized with eigenenergies $E_{\pm} = \pm \sqrt{( \tilde{J}(0)m + h_{\parallel} )^2 + \Gamma^2 } $ and eigenvectors
\begin{equation*}
 | \phi_+ \rangle = \begin{pmatrix} \cos{\frac{\theta}{2}} \\  \sin{\frac{\theta}{2}} \end{pmatrix}, 
 \quad | \phi_{-} \rangle = \begin{pmatrix} -\sin{\frac{\theta}{2}} \\ \cos{\frac{\theta}{2}} \end{pmatrix}
\end{equation*} 
with 
\begin{equation*}
\tan{\theta} = \frac{2\, \Gamma}{2 (\tilde{J}(0) m + h_{\parallel} )}, 
\end{equation*}  
so that the self-consistency condition for the thermal average (with $\beta = 1/k_B T$) is given by
\begin{align}
 m &= \langle \hat{S}^z \rangle = \frac{\langle \phi_{-} | \hat{S}^z | \phi_{-} \rangle e^{-\beta E_{-}}  + \langle \phi_{+} | \hat{S}^z | \phi_{+} \rangle e^{-\beta E_{+}} } 
				       { e^{-\beta E_{-}} + e^{-\beta E_{+}} } \notag \\
			       &=  \frac{ \tilde{J}(0) m + h_{\parallel} }{ \sqrt{\Gamma^2 + ( \tilde{J}(0) m + h_{\parallel})^2} } \tanh \left( \beta \sqrt{\Gamma^2 + ( \tilde{J}(0) m + h_{\parallel})^2 } \right)			       
\label{eq:mean_field_eq}
\end{align}

\begin{figure}
 \includegraphics[width=1.0\linewidth]{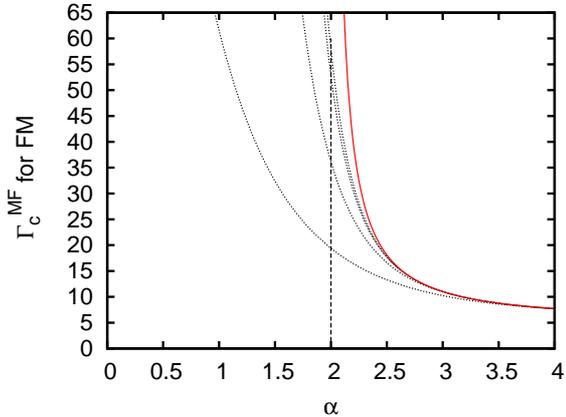}
 \caption{Mean-field phase boundary (red line) of the ferromagnetic Ising model on the triangular 
 lattice and in a transverse field as given by Eq.\eqref{eq:MF_phase_bounday}. The dotted lines are for 
 successively larger hexagonally shaped systems of radius $R=10,100,1000,$ and $2000$.}
 \label{fig:MF_phase_diagram_FM} 
\end{figure}

\begin{figure}
 \includegraphics[width=1.0\linewidth]{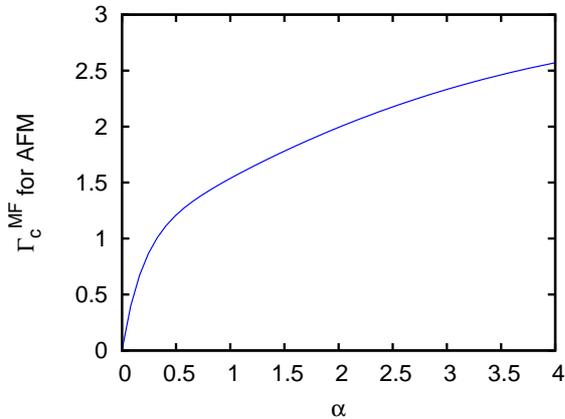}
 \caption{Mean-field phase boundary of the antiferromagnetic 
 Ising model on the triangular lattice and in a transverse field as given by Eq.\eqref{eq:MF_phase_bounday}.}
 \label{fig:MF_phase_diagram_AFM}
\end{figure}

In the limit $\Gamma \rightarrow 0 $ this reduces to the well-known mean-field equation for an Ising ferromagnet in a longitudinal field.
In the following we consider $h_{\parallel} = 0$ so that the only contribution to the longitudinal field is the mean field coming from the interaction
with other spins. Taking the limit $m \rightarrow 0$ gives the phase boundary $\frac{\Gamma}{\tilde{J}(0)} = \tanh(\beta_c \Gamma)$ with a 
zero-temperature critical point at $\Gamma_c = \tilde{J}(0)$ and a critical point a zero field at $k_B T_c = \tilde{J}(0)$.
At $T=0$ the order parameter increases as $ m = \pm \sqrt{1 - \frac{\Gamma}{\tilde{J}(0)}} $, as is typical of mean-field solutions.
Thus, in this admittedly very simple model the extent of the ordered phase scales with the interaction sum $\tilde{J}(0)$, which - as will be shown below - provides already qualitatively correct 
predictions of the phase boundary for both ferromagnetic and antiferromagnetic long-range interactions. 
For a mean-field treatment of the antiferromagnet we assume that the magnetic unit cell consists of three sites.
Then, the self-consistency equation for each of the sublattices A,B and C reads
{\footnotesize
\begin{equation}
 m_A \equiv \langle S_l^z \rangle = \frac{H_l^z(m_A, m_B, m_C, h_{\parallel})}{\sqrt{\Gamma^2 + \left( H_l^z \right)^2}} \tanh \left(\beta \sqrt{\Gamma^2 + \left( H_l^z \right)^2 }\right)
\end{equation}
}
with the longitudinal mean field $H_l^z = \tilde{J}_{AA}(0) m_A + \tilde{J}_{AB}(0)(m_B + m_C) $ acting at sites $l \in A$. The equations for $m_A$ and $m_B$
are obtained by cyclic permutation of the indices $A,B,C$. The interaction sum $\tilde{J}_{AA}(0) = \sum_{i \in A} \frac{J}{r_i^{\alpha}}$ gives the 
interaction of a spin in sublattice $A$ with all other spins in the same sublattice.
$\tilde{J}_{AB}(0) = \tilde{J}_{AC}(0) = \sum_{i \in B } \frac{J}{r_i^{\alpha}}$ is the interaction of a spin in sublattice $A$
with all spins in either of the other two sublattices $B$ or $C$, and $\tilde{J}(0) = \tilde{J}_{AA}(0) + \tilde{J}_{AB}(0) + \tilde{J}_{AC}(0)$. With the reasonable assumption
that the total magnetization has to be zero, $m_A + m_B + m_C = 0$,
the mean field simplifies to $H_l^z = m_A (\tilde{J}_{AA}(0) - \tilde{J}_{AB}(0))$. 
Thus, the mean-field phase boundary of the antiferromagnet is the same as that of the ferromagnet, except that it is scaled
by the interaction sum $\tilde{J}_{AA}(0) - \tilde{J}_{AB}(0)$.
The lattice sums are absolutely summable for $\alpha > 2$ (FM interactions) and $\alpha > 0$ (AFM interactions) and we compute them as a function of $\alpha$ 
analytically in the thermodynamic limit\footnote{ $\tilde{J}_{AA}(0;\alpha) - \tilde{J}_{AB}(0;\alpha) \equiv H_2(\alpha)$ is equivalent to the hexagonal lattice sum, i.e. the interaction sum of an
antiferromagnetic configuration on the (bipartite) hexagonal lattice which in turn corresponds to a sublattice magnetization $(1,-1,0)$ on the 
triangular lattice. 
A useful parametrization for the hexagonal lattice sum is \cite{Borwein2013}
\begin{multline*}
 H_2(2s = \alpha) = \\
 \frac{4}{3} \sum_{m,n= -\infty}^{\infty} \frac{\sin(n+1)\theta \sin (m+1)\theta - \sin n \theta \sin (m-1) \theta}{\left[ (n + \frac{1}{2}m)^2 + 3 (\frac{1}{2} m)^2 \right]^s }.
\end{multline*}
with $\theta = 2 \pi / 3$.
An analytical formula is given in \cite{Borwein2013} as $H_2(\alpha) = 3 (3^{1-\frac{\alpha}{2}} - 1) \zeta(\frac{\alpha}{2}) L_{-3}(\frac{\alpha}{2})$ 
in terms of the Riemann $\zeta$-function $\zeta(s) = \sum_{n=1}^{\infty} \frac{1}{n^s}$ and a Dirichlet L-series 
$L_{-3}(s) = 1 - 2^{-s} + 4^{-s} - 5^{-s} + 7^{-s} - 8^{-s} \cdots$. 
The exact result for the triangular lattice sum is \cite{Borwein2013}
\begin{align*}
 \tilde{J}(0;2s = \alpha)  &= \sum_{m,n=-\infty}^{\infty} \frac{1}{[n^2 + nm + m^2]^s} \\
                           &= 6\,\zeta(\frac{\alpha}{2}) L_{-3}(\frac{\alpha}{2}).
\end{align*}
}
,and numerically for a large triangular lattice with hexagonal boundaries.
Figs. \ref{fig:MF_phase_diagram_FM} and \ref{fig:MF_phase_diagram_AFM} show the resulting mean-field phase boundaries 
as a function of the decay exponent $\alpha$:
\begin{equation}
 \Gamma_c^{\text{MF}}(0; \alpha) = 
 \left\{
	\begin{array}{ll}
       \tilde{J}(0;\alpha) \\
       \tilde{J}_{\text{AA}}(0; \alpha) - \tilde{J}_{\text{AB}}(0;\alpha),     
       	\end{array} \right.
       	\label{eq:MF_phase_bounday}
\end{equation}
where the first line applies to FM and the second line to AFM interactions.
 To illustrate the issue of convergence with system size, the 
 lattice sum $\Gamma_c^{\text{MF}}(0; \alpha) = \tilde{J}(0;\alpha)$  was computed for a system of lattice points 
 with hexagonal shape, the size of which is parametrized by its radius $R$ (see Appendix).
 For $\alpha<2$, the lattice sum diverges as the system size tends to infinity; for $\alpha=2$
 it diverges logarithmically with the linear extent $R$ of the system. This can be seen from Fig.\ref{fig:MF_phase_diagram_FM}, where the dotted lines 
 correspond to systems with radius $R=10,100,1000$, and $2000$ (in units of the lattice constant).
 The curve of the phase boundary in the thermodynamic limit (red line) is 
 the exact result $\tilde{J}(0;\alpha) = 6 \, \zeta(\frac{\alpha}{2}) L_{-3}(\frac{\alpha}{2})$.
 

If the Ising interactions do not decay with distance, i.e. $\alpha=0$ and  $J_{ij} = J$ for all $(i,j)$, 
the Hamiltonian can be rewritten in terms of a single macroscopic spin operator $\hat{S}_{\text{tot}}^{z(x)} = \sum_{i=1}^N \hat{S}_{i}^{z(x)}$:
\begin{align}
 H &= \frac{J}{2} \left( \sum_{i=1}^N \hat{S}_i^z \right) \left( \sum_{j=1}^N \hat{S}_j^z \right) - \Gamma \left( \sum_{i=1}^N \hat{S}_i^x \right) \notag \\
   &= \frac{J}{2} (\hat{S}_{\text{tot}}^z)^2 - \Gamma \hat{S}_{\text{tot}}^x.
 \label{eq:LMG}
\end{align}
where we have omitted an additive constant. This is an anisotropic variant of the Lipkin-Meshkov-Glick model \cite{Ribeiro2008}.
For this infinitely coordinated Ising model the lattice structure becomes irrelevant. 
The AFM model ($J>0$) for $\Gamma = 0$ has a degeneracy which is exponential in $(N/2)$: Any configuration with $S_{\text{tot}}^z = 0$ is a ground state.
The transition into
the paramagnetic state should occur immediately at $\Gamma_c = 0^{+}$ since for $S_{\text{tot}}^{z} = 0$ there is no opposing Ising interaction.

As for nearest neighbour interactions, the Fourier transformation of $J_{ij} = J/|{\bf r}_i - {\bf r}_j|^{\alpha}$ displays minima at the corners 
of the hexagonal Brillouin zone. However, they become increasingly shallow as $\alpha$ decreases and vanish around $\alpha_c \approx 1$.
This means that for $\alpha < \alpha_c$ the lattice structure cannot dictate the ordered state to be selected by the quantum fluctuations.

\section{The QMC Method}
In Ref. \cite{Sandvik2003} Sandvik proposed 
a Stochastic Series Expansion (SSE) QMC method that can 
deal with transverse field Ising models with arbitrary, long-range or frustrated, interactions. In the case
of long-range interactions, it avoids the interaction summation that is typically necessary, and 
the scaling of the CPU time with system size is reduced from $N^2$ to $N \ln(N)$. 
The main trick consists in adding constants $|J_{ij}|$ to the Ising bond operators
\begin{equation} 
 H_{i,j} =  | J_{ij} | - J_{ij}\sigma_i^z \sigma_j^z, \quad i \neq j
 \label{eq:SSE_weights}
\end{equation}
in such a way that only satisfied bonds (i.e. FM bonds, $J_{ij} < 0$, on FM spin configurations, $\sigma_i^z = \sigma_j^z$, or AFM
bonds on AFM spin configurations) have non-zero weight. This constraint, which is active in the propagation direction
of the SSE algorithm, obviates techniques such as Ewald summation of the interactions in real space that are typically used for 
long-range interacting systems.
Furthermore it is evident from \eqref{eq:SSE_weights} that even a frustrated 
transverse field Ising model has no sign problem in the $\sigma^z$ basis since any negative matrix elements can be shifted 
by a constant. 
As the Monte Carlo update depends crucially on the presence of single-spin flip (transverse field) terms in the Hamiltonian, ergodicity 
may be lost
for small transverse fields. By the same token, we find that the algorithm does not perform well at finite temperatures as soon as 
the thermal fluctuations become comparable to the quantum fluctuations,  $k_B T \sim \Gamma$. On the other hand, this update mechanism
proves extremely efficient in the case of large transverse fields.

Since the Ising model in a transverse field does not contain
off-diagonal two-body operators, it is not possible to use a loop update. Instead, we employ a 
multi-branch cluster update \cite{Sandvik2003}. The simulation cell with open boundaries is hexagonal and parametrized by its ''radius`` $R$ so that the total number of 
spins is $N(R) = 1 + 3 R(R+1)$. To be closer to the experimental situation in a cylindrically symmetric trapping potential, for 
$R \ge 7$ additional spins were included which 
extend the hexagon to an approximately circular shape (see Fig. \ref{fig:simcell} in the Appendix)
so that in those cases the number of spins is larger than given by this formula. Due to the six-fold symmetry the total number of spins is always $N=6n+1$ where 
$n$ is an integer.

\emph{Simulation parameters:}
While for FM interactions around $50 - 100$ thermalization steps and $10^4$ measurement steps were sufficient 
to study the critical behaviour, in the long-range, frustrated AFM case $\approx 10^4$ thermalization steps
and up to $4 \cdot 10^6$ measurement steps were required due to long thermalization and autocorrelation times. 
The largest systems studied consisted of $613$ spins for FM interactions and $301$ spins for AFM interactions, respectively.
The simulations were performed at $T=1/(2R)^z$, where $2R$ is the diameter of the simulation cell and $z=1$ 
is the dynamical critical exponent. This choice ensures that the thermal energy $k_B T$ is smaller than
the finite-size gap such that the system is expected to be at a temperature that is effectively $T=0$. On the other hand, rescaling 
the temperature with system size according to this formula is generally necessary in order to obtain suitable 
data sets for finite-size scaling (see discussion below).

\emph{Code verification:}
The Monte Carlo code was sucessfully checked against Lanczos exact diagonalization (ED) on a hexagonal 
system of radius $R=2$ with $N=19$ spins. 
\begin{figure}
 \includegraphics[width=1.0\linewidth]{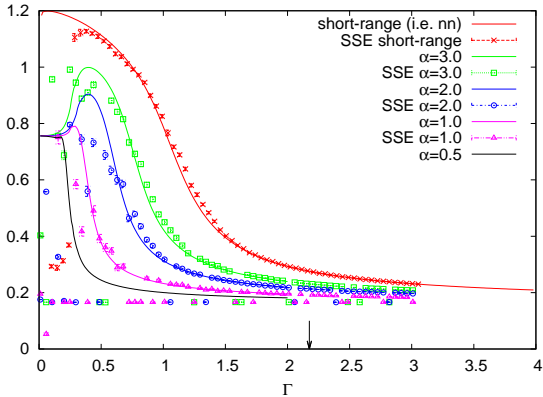}
 \caption{Comparison between Lanczos exact diagonalization and SSE 
 QMC for a hexagonal system of 19 spins with AFM interactions. Shown are the fluctuations
 of the modulus of the complex $XY$ order parameter $\langle m^2 \rangle$. In the region of interest around $\Gamma \approx 1$ the results from both algorithms agree nicely. For some values of $\Gamma$ the mean fluctuation of the magnetization could not be determined due to technical issues. 
 Those points lie around $\langle m^2 \rangle = 0.2$ and are meaningless. 
 For comparison, the critical value in the AFM mean field theory for $\alpha=2.5$ is indicated by an arrow at $\Gamma_c^{\text{MF}} (\alpha=2.5)=2.1758$.
 }
 \label{fig:Lanczos}
\end{figure}
Note that the next larger sytem size R=3, N=37 with \emph{open boundary conditions} was not amenable to exact diagonalization since momentum is not a good quantum number.
The agreement between QMC and ED for ferromagnetic interactions is excellent. 
For antiferromagnetic interactions, on the other hand, the error bars are only meaningful in the paramagnetic phase. 
In Fig. \ref{fig:Lanczos} the square of the 
clock order parameter \eqref{eq:COparameter}, that is its fluctuations, 
versus the transverse field $\Gamma$ is shown for different decay powers $\alpha$.
There is good agreement between QMC and ED in the paramagnetic phase whereas in the ordered phase the curves do not agree
within errorbars. This is due to the fact that in the ordered phase the fluctuations are non-Gaussian, exhibiting asymmetric tails as
can be seen from the histogram of the squared clock order parameter in Fig. \ref{fig:histograms} in the Appendix.
Furthermore for small transverse fields ergodicity may be lost when the quantum clusters 
used in the Monte Carlo update percolate in imaginary time (see Ref. \cite{Sandvik2003} for a discussion of this issue).
As a consequence, the error bars in our simulations of the antiferromagnet are only reliable for the paramagnetic phase and for the onset of order,
but not within the ordered phase. For determining the critical field only the former parameter regions are needed so that 
the phase boundary can still be obtained quantitatively. 
As an illustration of the quantitative inaccuracy of mean field theory, an arrow has been added in Fig. \ref{fig:Lanczos}
indicating the critical value in mean field theory for $\alpha=2.5$, which lies at $\Gamma_c^{\text{MF}} (\alpha=2.5)=2.1758$.

\section{Results}

\subsection{Long-range ferromagnet on the triangular lattice}
The value of the critical field can be estimated with various methods.

\begin{figure}
\includegraphics[width=1.0\linewidth]{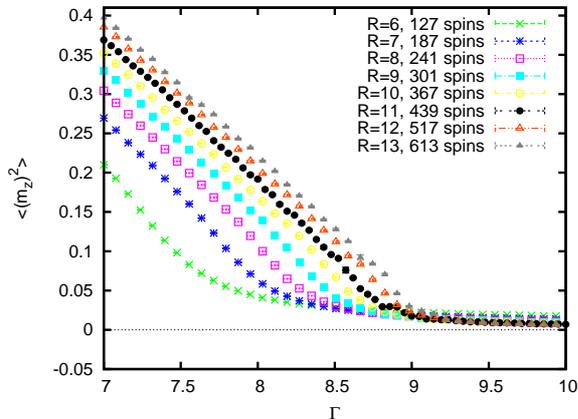}
 \caption{Squared magnetization per site for \mbox{$\alpha=3.0$}. For a mean-field transition, $\beta = 0.5$ and $\langle m_z^2\rangle$ should vanish 
 linearly in the vicinity of the critical field.}
 \label{fig:FM_alpha3_magnzz}
\end{figure}

We determine the critical field by locating the crossing points 
of the Binder cumulant \cite{Binder1981} for consecutive system sizes and extrapolating to the thermodynamic limit.
For a scalar order parameter the appropriate Binder cumulant is
\begin{equation}
 U_L = \frac{3}{2}\left(1 - \frac{\langle m^4 \rangle}{3 \langle m^2 \rangle ^2}\right), 
\end{equation}
where $\langle m^{i} \rangle$ is the $i$-th moment of the $z$-component of the magnetization (see Fig. \ref{fig:FM_alpha3_magnzz}).
It has scaling dimension zero since at the critical point the power laws in the linear system size $R$ for $\langle m^4 \rangle$
and $\langle m^2 \rangle$ cancel out. In the limit $R \rightarrow \infty$ the Binder cumulant has the following 
properties: $U_R \rightarrow 1$ in the ordered phase, $U_R \rightarrow 0$ in the disordered phase and 
at the critical point $U_R \rightarrow U^{\star}$, i.e. the Binder cumulants for different system 
sizes intersect at a common point $U^{\star}$, which is also universal.
Often there are subleading finite-size corrections so that the crossing points for pairs of system sizes
drift providing a size-dependent critical point which typically converges much faster than the usual finite-size 
shift $\propto R^{-1/\nu}$
of other quantities with singular behaviour at the critical point \cite{Sandvik2010}.

\begin{figure}
\includegraphics[width=1.0\linewidth]{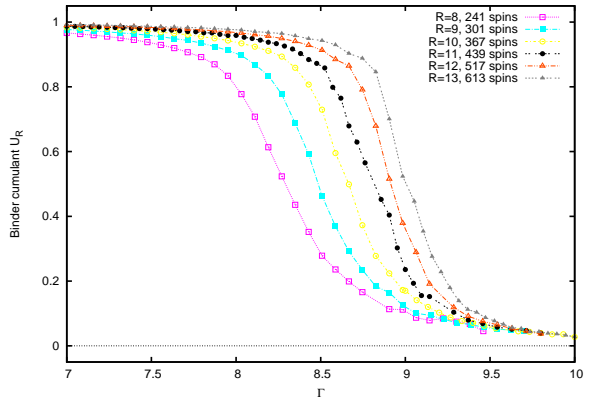}
\caption{Binder cumulants for $\alpha=3.0$.}
\label{fig:FM_alpha3_Binder}
\end{figure}

\begin{figure}
\includegraphics[width=1.0\linewidth]{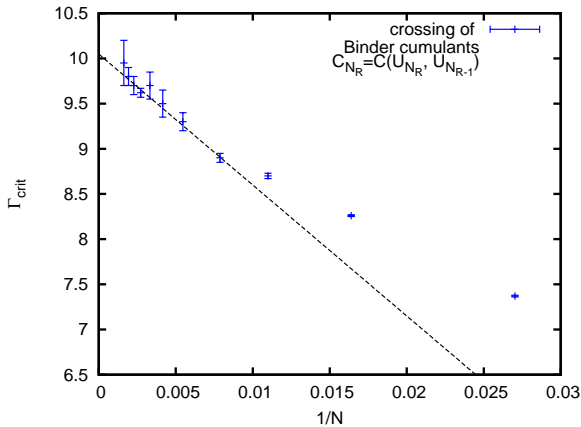}
  \caption{Extrapolation of the crossing points of the Binder cumulant for $\alpha=3.0$. $N_R$ is the number of lattice
  sites for a system with radius $R$; $C_{N_R} = C(U_{N_R}, U_{N_{R-1}})$ denotes the value of the Binder cumulant $U_{N_R}$ for system size
  $N_R$ at the crossing point with the Binder cumulant for the sucessively smaller system size $N_{R-1}$. The range of system sizes is
  $R=3,4,\ldots,13$.}
    \label{fig:Binder_FM}
\end{figure}

Fig. \ref{fig:FM_alpha3_Binder} shows the Binder cumulant $U_R$ for system sizes $N = 241, 301, 367, 439, 517, 613 $.
The fact that the crossing points of the Binder cumulants for successive system sizes occur at small values of $U_R$ renders 
this otherwise very accurate method of determining the critical field problematic. (An error propagation shows that small $\langle m^2 \rangle$ lets the error 
on the Binder cumulant increase drastically.)  \mbox{Fig. \ref{fig:Binder_FM}}
shows the extrapolation of the crossing points of the Binder cumulant vs. $1/N$ 
which leads to an estimate for the critical field of $\Gamma_c \approx 10.0$.
Smaller system sizes are more affected by boundary effects due to the open boundary conditions and 
therefore they were excluded from the fit. The fact that a fit including only the smaller system sizes would underestimate
the critical transverse field is consistent, as a larger boundary to bulk ratio reduces the restoring Ising interaction energy
relative to the energy of the spins aligned with the transverse field, the latter being independent of the boundary to bulk ratio.

\begin{figure}
\includegraphics[width=1.0\linewidth]{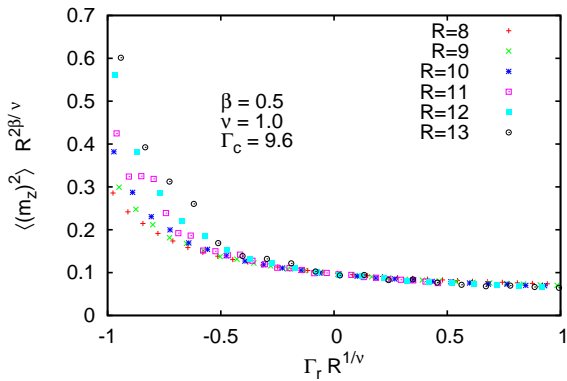}
\caption{Data collapse for $\alpha=3.0$ with mean-field critical exponents $\beta=1/2$, $\nu = 1$. The linear extent of the system is measured in terms 
of the radius $R$ of the simulation cell. $\Gamma_r = (\Gamma - \Gamma_c)/\Gamma_c$ denotes the reduced field. The best data collapse is achieved
for $\Gamma_c \approx (9.6 \pm 0.1)J$.}
\label{fig:FSS_scaling}
\end{figure}
The critical behaviour of long-range ferromagnetic quantum Ising and rotor models has been studied by Dutta and Bhattacharjee \cite{Dutta2001}
using field theory and renormalization group equations. They find that in two dimensions the critical exponents attain values of the short-range system
for $\alpha \ge 4$. For $\alpha_{u} \le \alpha < 4$ the critical exponents depend continuously on $\alpha$ and reach mean-field values for
$\alpha < \alpha_{u}$ where $\alpha_u = 10/3$ is the upper critical range in two dimensions. We verify this prediction of mean-field critical exponents 
for $\alpha=3$ using finite-size scaling.
According to the finite-size scaling hypothesis\cite{Cardy}, close to the critical point the order parameter
squared scales with the linear system size $R$ and the reduced control parameter $\Gamma_r = \frac{\Gamma - \Gamma_c}{\Gamma_c}$ as
$\langle m^2 \rangle (\Gamma_r,T,R) = R^{-2\beta/\nu} g(\Gamma_r R^{1/\nu},TL^z)$ where $\nu$ is the critical exponent for the correlation length,
$\xi \propto |\Gamma_r|^{-\nu}$, $\beta$ is the exponent for the order paramenter, $\langle m^2 \rangle \propto |\Gamma_r|^{2 \beta}$
and $z=1$ is the dynamic critical exponent. If the temperature is rescaled with system size according to $T=1/(2R)^{z}$,
the second argument of the scaling function $g$ becomes a constant for all data sets, and then the scaling function depends only on a single parameter. 
Then, when plotting $y_R = \langle m^2 \rangle (\Gamma_r,R) R^{2\beta/\nu} $ against $x_R = \Gamma_r R^{1/\nu}$,
data sets for different linear system size should collapse onto the scaling function $g(x)$
if the critical exponents $\beta$ and $\nu$ are chosen appropriately. 

It is a well-established fact that the 2d transverse-field Ising model on any integer-dimensional regular lattice
has a dynamical critical exponent of $z=1$ \cite{Sachdev}. 
It is argued in Ref. \cite{Dutta2001} that the dynamical exponent for the long-range transverse-field Ising model 
depends continuously on the decay exponent $\alpha$, reaching $z=1$ only for $\alpha=4$ (or for $\sigma=2$ in their
notation, where $\alpha=d+\sigma$ and the spatial dimensionality is $d=2$). For $\alpha<4$, Ref. \cite{Dutta2001} predicts $z<1$, which reflects
the expectation that the correlation length in imaginary time grows slower than that in the spatial direction 
as a consequence of the long-range interactions.
On the other hand, in \cite{Baek2011} it was found by extensive QMC simulations that the infinitely coordinated 
Ising model in a transverse field has $z=1$. We did not attempt to determine $z$ numerically, but naively assumed $z=1$ for rescaling 
the temperature with system size. This should not affect the validity of the finite size scaling
as long as the temperature is always below the respective finite-size gap such that the system is effectively at $T=0$.
Then, the second variable $TL^{z}$ in the scaling function drops out.

Fig. \ref{fig:FSS_scaling} shows
that for $\alpha=3$ a satisfactory data collapse can be achieved with mean-field critical exponents $\beta=\frac{1}{2}$ and $\nu=1$.
The best data collapse is obtained for $\Gamma_c = (9.6 \pm 0.1)J$. 
A critical field of $\Gamma_c = (10.0 \pm 0.4)J$ is thus consistent
with the two estimates based on the extrapolation of the crossing points of the 
Binder cumulant and the data collapse.
This value is, as expected, smaller but very close to the value $\Gamma_c^{\text{MF}} (\alpha=3.0) = 10.95$ of mean field theory, 
which is not obvious since the critical field is not a universal quantity.
Since for increasingly long-range interactions mean field theory should become a better approximation,
with the true phase boundary never exceeding that given by mean field theory, this fact implies
that for $\alpha < 3$ the mean-field phase boundary gradually becomes the \emph{true} phase boundary.

The quantum critical behaviour of the dipolar $(\alpha=3)$ Ising ferromagnet differs in a subtle way from the behaviour at the finite-temperature phase transition
in zero field: 
While in $d=2$ the exponent $\alpha=3$ places the classical system in its thermal phase transition 
on the boundary between long-range (i.e. $\alpha$-dependent) and mean-field critical exponents, leading to logarithmic corrections for the divergence of the correlation
length and susceptibility at the critical point \cite{Fisher1972}, the increased dimensionality $d_{\text{eff}}=d+z$
of the quantum critical point results in mean-field critical behaviour \emph{without any corrections}. 
Therefore one can expect that for the dipolar Ising FM an experiment 
in the thermodynamic limit could show that the finite-temperature phase transition has logarithmic corrections to its critical behaviour 
whereas the zero-temperature phase transition does not.
However, it is extremely difficult to observe the presence or absence of logarithmic corrections in numerical studies on finite-size systems.
A similar dimensional cross-over of the values of critical exponents has been experimentally observed in the 3d nearest neighbour Ising model which exhibits 
mean-field critical exponents at its quantum critical point in a transverse field \cite{Erkelens1986}. Of course, this is a much more pronounced effect than the absence or presence of logarithmic 
corrections in the dipolar ferromagnet.

\begin{figure}
 \includegraphics[width=1.0\linewidth]{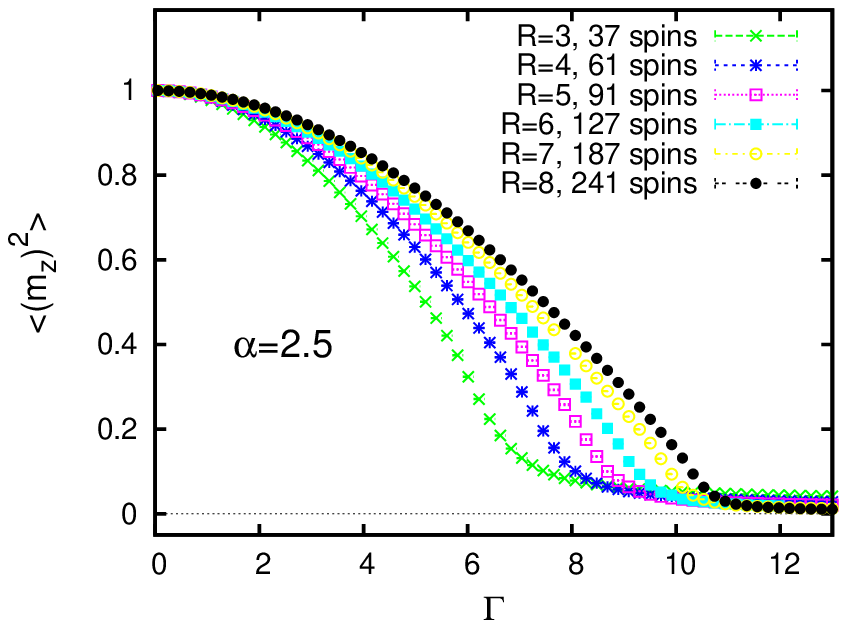}
 \includegraphics[width=1.0\linewidth]{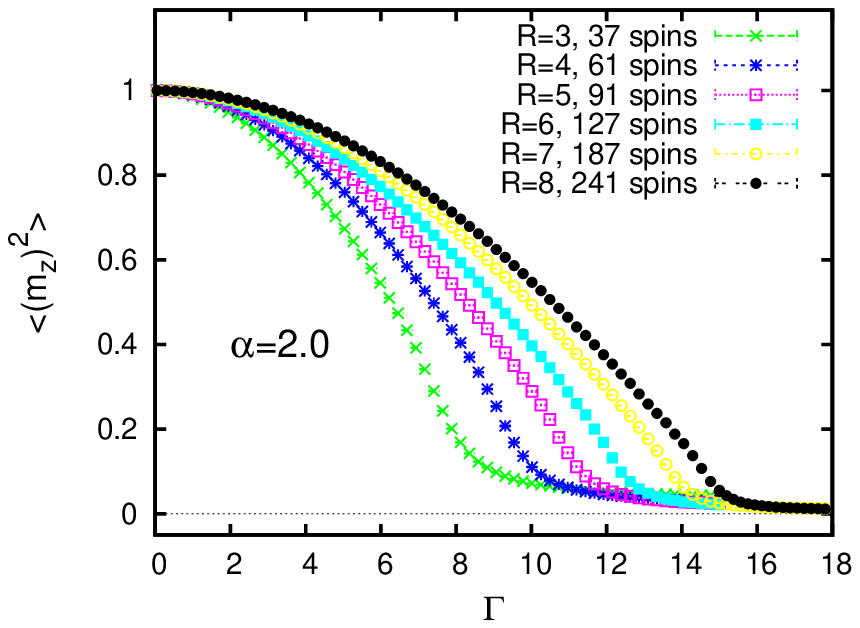}
 \includegraphics[width=1.0\linewidth]{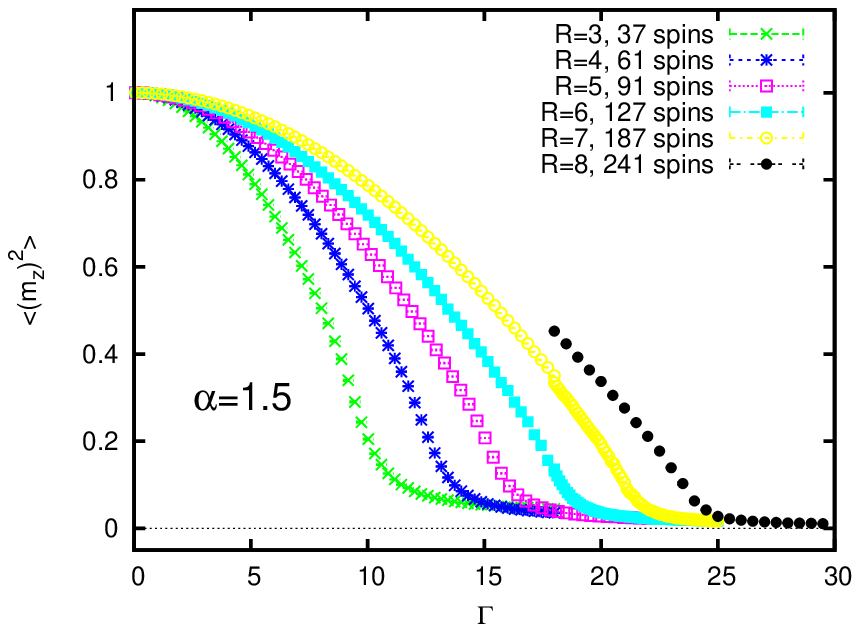}
 \caption{Squared magnetization per site for \mbox{$\alpha = 2.5, 2.0$, and $1.5$}. The errorbars are smaller than the symbol size.}
 \label{fig:FM_alphax_magnzz}
\end{figure}

Fig. \ref{fig:FM_alphax_magnzz} shows magnetization curves for $\alpha = 2.5, 2.0$ and $1.5$. In view
of possible experiments on finite-size systems it needs to be pointed out that the approach of the critical
field to the value of mean field theory in the thermodynamic limit is very slow for $\alpha < 3$. 
For $\alpha \le 2$ the interaction energy $\tilde{J}(0)$ is superextensive in the system size and the mean-field 
critical field diverges in the thermodynamic limit. This is typically remedied by introducing a regularizing factor $1/N$ into the model,
which corresponds to rescaling the energy and time scales and gives a well-defined thermodynamic limit.
Since the experimental system under consideration is always finite with a few hundreds to thousand spins,
the question of stability in the thermodynamic limit is not important.

\subsection{Long-range antiferromagnet on the triangular lattice}

\emph{Clock order parameter:}
The complex $XY$ order parameter following from the Landau-Ginzburg-Wilson analysis 
can be written as (see \cite{Isakov2003} and references therein):
\begin{equation}
 m e^{i \theta} \equiv ( m_A + m_B e^{i(4 \pi /3)} + m_C e^{i(-4 \pi /3)} ) / \mathcal{C}
 \label{eq:COparameter}
\end{equation}
where  $m_j, j=A,B,C$ are the magnetizations of the three sublattices.
The normalization is $\mathcal{C}=\sqrt{3}$ for $(1,-1,0)$ order and $\mathcal{C} = 3/2$ for $(1, -\frac{1}{2}, -\frac{1}{2})$ order.

\emph{Binder cumulant:}
The correct Binder cumulant for an n-component order parameter (n=2 in our case) is \cite{Sandvik2010}
\begin{equation}
 U = \frac{n+2}{2} \left( 1 - \frac{n}{n+2} \frac{\langle m^4 \rangle}{\langle m^2 \rangle ^2} \right)
\end{equation}

Whether the Binder cumulant has to be extrapolated against $1/N$ or $1/N^{x}$
where $x$ is another power greater than $1$ depends on the subleading finite-size
corrections of the Binder cumulant\cite{Binder1981}. The values for $\Gamma_c$ in the thermodynamic limit vary strongly 
with the chosen extrapolation scheme, but here we attempt to determine $\Gamma_c$ for large, but finite system sizes. 

\begin{figure}
 \includegraphics[width=0.8\linewidth]{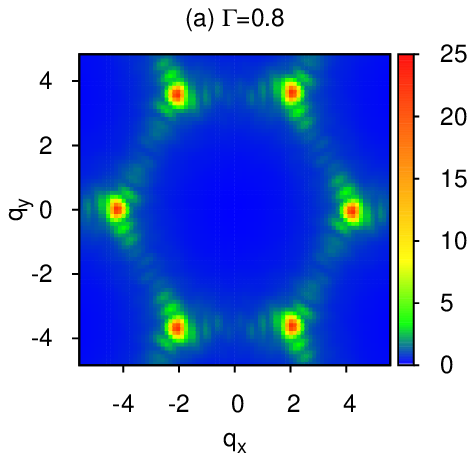}
 \includegraphics[width=0.8\linewidth]{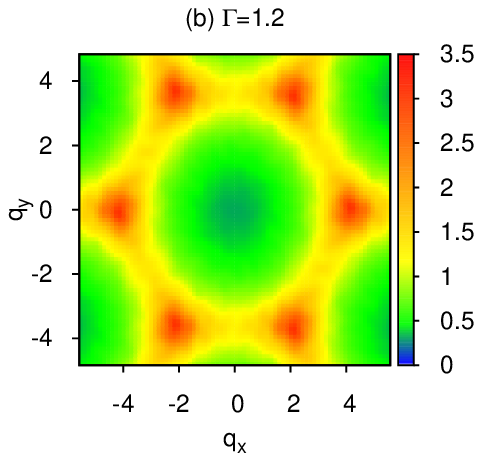}
 \includegraphics[width=0.8\linewidth]{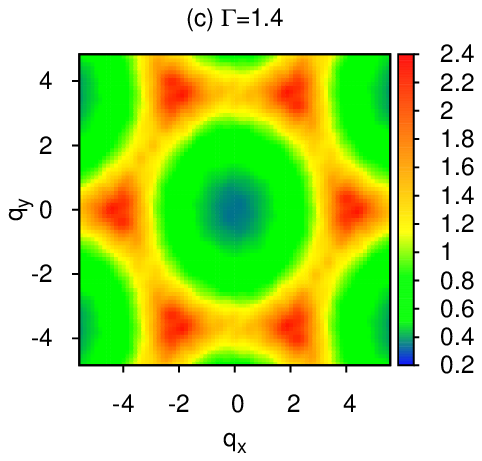}
 \caption{Structure factor for $R=8, N=241$ spins and $\alpha=3.0$ for different values of the transverse field $\Gamma$ below
($\Gamma = 0.8J$) and above ($\Gamma = 1.2J, 1.4J$) the phase transition, i.e. in the clock ordered and the fully x-polarized phase, respectively.}
\label{fig:sqzz_AFM}
\end{figure}

\emph{Structure factor:} The stucture factor defined as 
\begin{equation*}
 S({\bf q}) = \langle S_{{\bf q}}^z S_{-{\bf q}}^z \rangle = \frac{1}{N} \sum_{ij} \langle S_i^z S_j^z \rangle \, e^{i {\bf q} \cdot ({\bf r}_i - {\bf r}_j)}
\end{equation*}
diverges $\sim N$ if there is order with wavevector ${\bf q} = {\bf Q}$. Fig. \ref{fig:sqzz_AFM} shows the structure factor for $R=8$, $N=241$ spins at transverse fields
$\Gamma$ below ($\Gamma=0.8$) and above ($\Gamma=1.2$ and $1.4$) the phase transition to a clock ordered phase. 
The Bragg peaks at ordering vectors ${\bf Q}_{\pm} = (\pm 4 \pi/3,0)$ and vectors related by reciprocal lattice vectors, i.e. at the corners of 
the hexagonal Brillouin zone, clearly indicate clock order without any trace of competing orderings. 

\begin{figure}
 \includegraphics[width=1.0\linewidth]{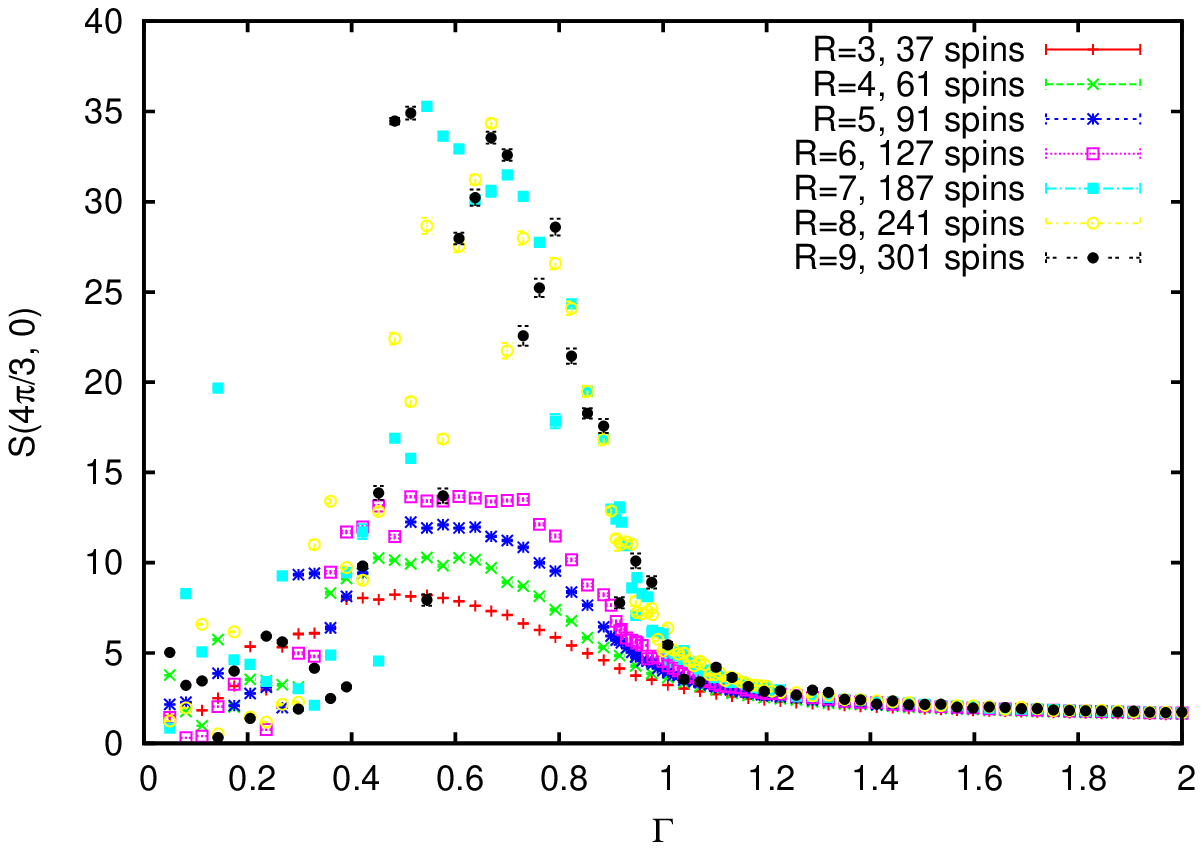}
 \caption{Structure factor $S({\bf Q})$ at \mbox{${\bf Q}_{+} = (4\pi/3, 0)$} for $\alpha=3$.}
 \label{fig:sqzz_dome}
\end{figure}

\begin{figure}
\includegraphics[width=1.0\linewidth]{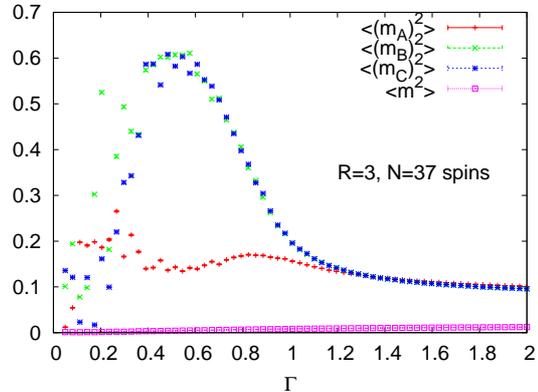}
\includegraphics[width=1.0\linewidth]{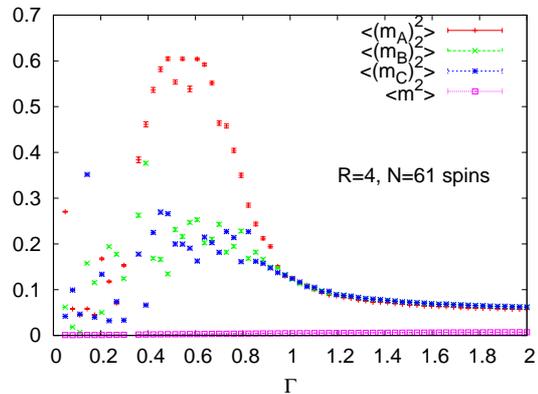}
\includegraphics[width=1.0\linewidth]{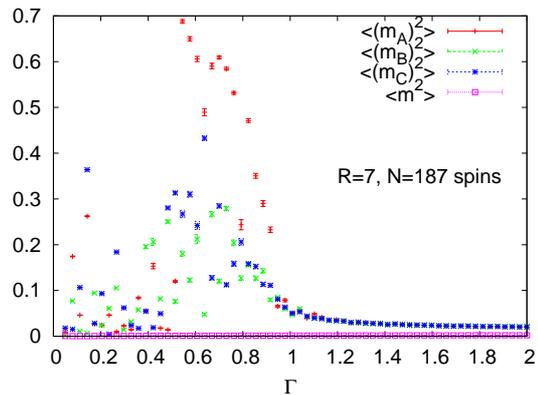}
 \caption{Squared sublattice magnetizations for decay exponent $\alpha=3.0$ for R=3, N=37 spins (upper panel);
  R=4, N=61 spins (middle panel);
  and R=7, N=183 spins (lower panel).}
 \label{fig:sublattice_magn}
\end{figure}

In Fig. \ref{fig:sqzz_dome} 
the structure factor at ${\bf Q}_{+}$ is presented for $\alpha=3.0$ and for different system sizes. The dome-like structure indicates a 
clock-ordered phase roughly between $\Gamma \approx 0.2$ and $\Gamma \approx 1.0$. As opposed to the nearest-neighbour AFM,
where clock order appears for infinitesimally small $\Gamma = O ^+$, in the long-range case there is a threshold in $\Gamma$ for the onset of clock order.
It was not possible to perform a scaling analysis of the height of the structure factor due to metastable states that appear for larger systems (see below):
Already for systems with radius $R=7,8,9$ the structure factor fails to increase further.
Both the ferrimagnetic $(1, -\frac{1}{2}, -\frac{1}{2})$ and the 
partially antiferromagnetic $(1, -1, 0)$ states have the same ordering vectors being distinguished only by the value of $\vartheta$ in Eq. \eqref{eq:LGW_Hamiltonian}.
However, while for the partially antiferromagnetic state the wavevectors ${\bf Q}_{+}$ and ${\bf Q}_{-}$ correspond to two different degenerate states,
for the ferrimagnetic state both wavevectors give the same spin configuration. Therefore, when spontaneous symmetry breaking occurs in the thermodynamic limit
the partially antiferromagnetic state will manifest itself in Bragg peaks at either ${\bf Q}_{+}$ (and equivalent corners of the first Brillouin zone)
or ${\bf Q}_{-}$. The ferrimagnetic state, on the other hand, will exhibit Bragg peaks at both ${\bf Q}_{+}$ and ${\bf Q}_{-}$, that is at all six
corners of the Brillouin zone. Since there is no spontaneous symmetry breaking in a finite system, Monte Carlo simulations - and also 
experiments on finite systems - do not allow to distinguish between 
the two types of states at the level of the structure factor, and 
one needs to look at the sublattice magnetization.
The squares of the sublattice magnetizations shown in Fig. \ref{fig:sublattice_magn} (middle panel)
for R=4, N=61 spins, $\alpha=3.0$, are consistent with the ferrimagnetic state $(M, -\frac{M}{2}, -\frac{M}{2})$ where $M$ is below the saturation value of 1. 
The sublattice magnetizations compensate each other 
so that there is no net magnetic moment. 
This justifies the assumption $m_A+m_B+m_C=0$ in the mean-field analysis of section \ref{sec:mean-field}.
For N=37 spins (Fig. \ref{fig:sublattice_magn}, upper panel) or N=61 spins (Fig. \ref{fig:sublattice_magn}, lower panel), 
one sublattice must have one spin more than
the other two, which explains the small deviation of $\langle(m_z)^2\rangle$ from zero.
In a larger system with N=187 spins (Fig. \ref{fig:sublattice_magn}, lower panel) where 
metastabilities are more pronounced (see Fig. \ref{fig:histograms}
in the Appendix) such that an interpretation is difficult, the ordered phase has sublattice magnetizations $(m_B+m_C, -m_B, -m_C)$ which would reduced 
to the ferrimagnetic state if $m_B \approx m_C = \frac{M}{2}$.
For a small system of $N=37$ spins (Fig. \ref{fig:sublattice_magn}, upper panel) the structure of the sublattice magnetizations
is quite different with two sublattices having a larger modulus of the magnetization than the third sublattice, $|m_A|, |m_B| > |m_C|$.
Apparently, resonance processes due to the transverse field which stabilize the clock ordered 
phase \cite{Moessner2001} and its characteristic sublattice magnetization cannot fully develop on this small 
system as they are strongly influenced by the open boundary conditions.

While the non-saturated magnetization \mbox{$M<1$} could be attributed 
to the difficulties of the Monte Carlo update in the ordered phase, it has been argued in Ref. \cite{Coppersmith1985} 
for the closely related stacked triangular short-range AFM that Landau-Ginzburg-Wilson theory is unreliable for the 
low-temperature behaviour and that the low-temperature phase should not be in 
the ferrimagnetic state $(1, -\frac{1}{2}, -\frac{1}{2})$. Based on entropy considerations, in \cite{Coppersmith1985}
a three sublattice structure was conjectured in which the spin chains in the stacking
direction are fully ordered and where most configurations are such that the chains on two sublattices 
align antiparallel while the third one is randomly oriented. This phase has been referred to as the 
3d analog of the 2d Wannier phase \cite{Heinonen1989, Wannier1950} and its character was subsequently 
supported by Monte Carlo simulations \cite{Heinonen1989, Kim1990}. 
More recent Monte Carlo simulations \cite{Zukovic2013} also favour this szenario
with sublattice magnetizations $(m_A, m_B, m_C) = (M,-M,0)$
whose magnitude $M$ is unsaturated, $M<1$, and decays as a power law with system size 
such that in the thermodynamic limit all sublattice 
magnetizations vanish and the low-temperature phase of the stacked triangular 
AFM shows no long-range order. 
In view of this controversy in a closely related system and 
because a reliable scaling analysis of the structure factor \emph{within} the order phase was not possible 
in our simulations due to metastable states,
a conclusive statement about true long-range order in the thermodynamic limit cannot be made. 

\begin{figure}
\includegraphics[width=1.0\linewidth]{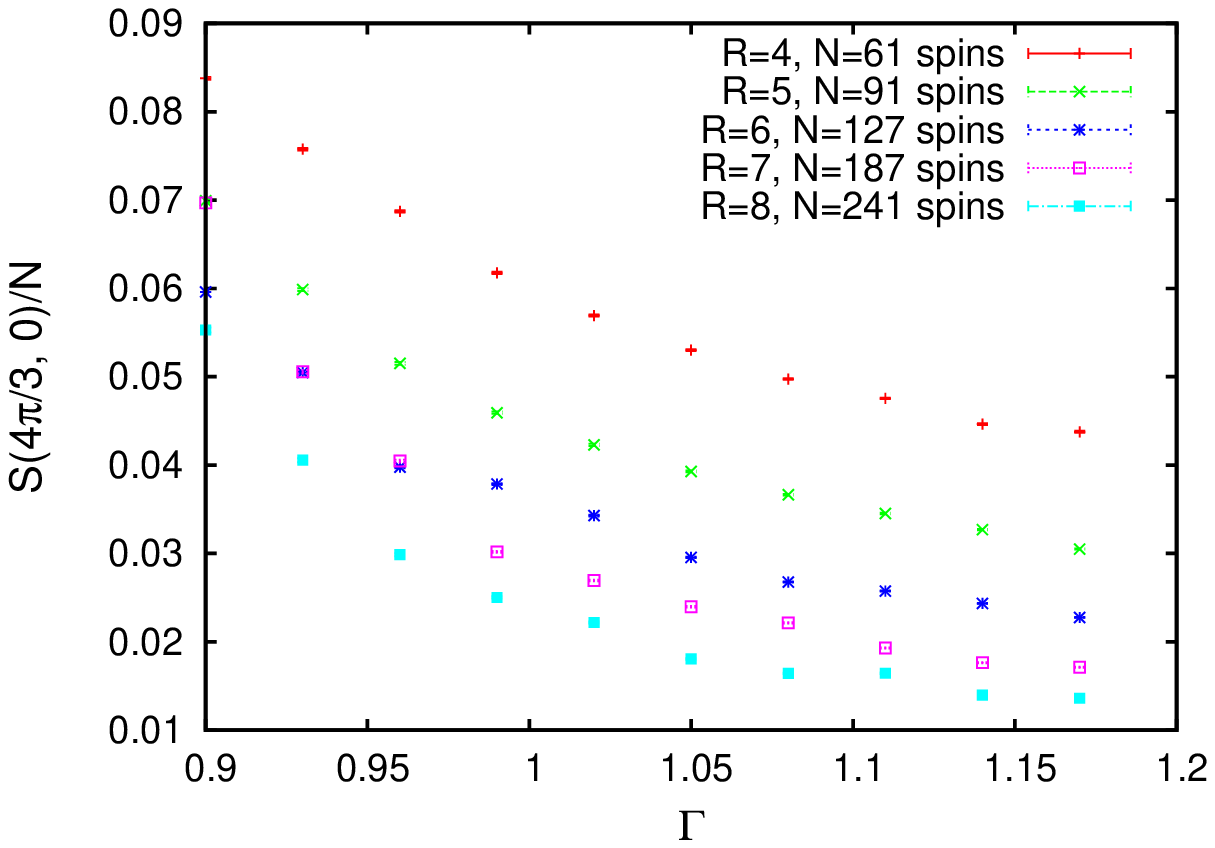}
\caption{Rescaled structure factor $S({\bf Q})/N$ at ${\bf Q}_{+} = (4 \pi /3,0)$ around the critical field $\Gamma_c = 1.05$ for $\alpha=3$. }
\label{fig:sqzz_rescaled}
\end{figure}

\begin{figure}
\includegraphics[width=1.0\linewidth]{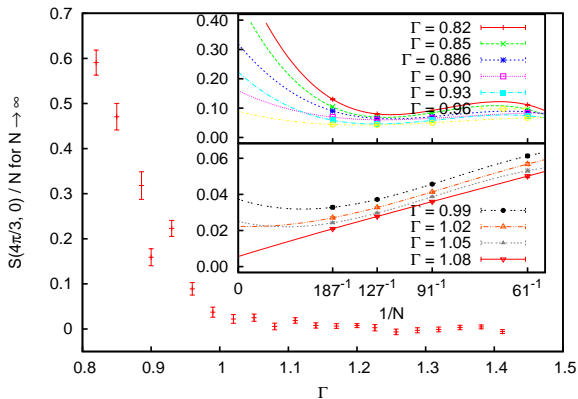}
\caption{Bootstrap analysis for decay exponent $\alpha=3$. The main panel shows the
 structure factor $S({\bf Q})/N$ at ${\bf Q}_{+} = (4 \pi /3,0)$, extrapolated to the thermodynamic limit. The
 extrapolation vs. $1/N$ with a third-order polynomial is shown in the inset for different transverse fields $\Gamma$. }
\label{fig:bootstrap_LR3AFM}
\end{figure}

\emph{Location of the quantum critical point:}
Fig. \ref{fig:sqzz_rescaled} shows the rescaled structure factor $S({\bf Q})/N$, where $N$ is the number of spins, at the ordering vector ${\bf Q}_{+} = (4\pi/3,0)$
around the critical field.
In order to determine the quantum critical point in the thermodynamic limit 
we extrapolate the rescaled structure factor $S({\bf Q}_{+})/N$ versus inverse system size $1/N$. 
The extrapolation is done with a third-order polynomial fit in $1/N$ as shown in the inset of Fig. \ref{fig:bootstrap_LR3AFM}.
The value in the thermodynamic limit  $1/N \rightarrow 0$ has an error attached to it which is determined 
via the bootstrapping method: The fitting procedure is repeated $10^5$ times adding Gaussian noise
to the data points with a spread of the size of the errorbars attached to the points. Then the distribution
of extrapolated values gives a mean value and quantifies its error. In this way we obtain the thermodynamic
limit of $S({\bf Q}_{+})/N$ in the main panel of Fig. \ref{fig:bootstrap_LR3AFM}.
From the field-dependence of the structure factor $S({\bf Q}_+)/N$ at ${\bf Q}_+ = (4\pi/3,0)$, extrapolated to the thermodynamic limit, 
we can make an estimate of the critical field $\Gamma_c = 1.00 \pm 0.05$ at which the structure factor vanishes.
Clearly, there is some arbitrariness in the choice of extrapolation scheme regarding the degree of the extrapolation polynomial
and the included system sizes. 
\begin{figure}
 \includegraphics[width=1.0\linewidth]{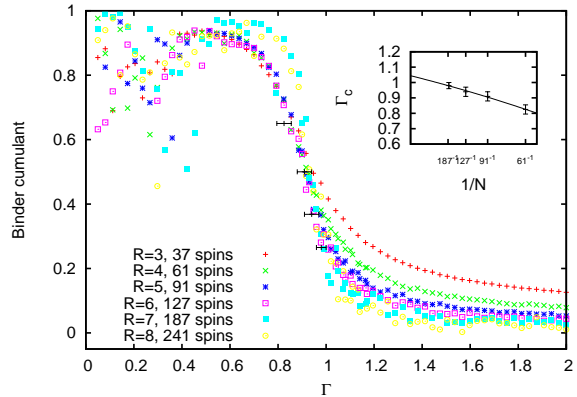}
 \caption{Binder cumulants $U_{N_R}$ for different system sizes; $\alpha=3.0$. Inset: Extrapolation of their crossing points $C_{N_R} = C(U_{N_R}, U_{N_{R-1}})$,
 which are indicated with horizontal error bars in the main graph, 
 versus inverse system size.}
 \label{fig:Binder_AFM_alpha3}
\end{figure}
From the crossing points of the Binder cumulant (see Fig.\ref{fig:Binder_AFM_alpha3})
a critical field of \mbox{$\Gamma  = 1.05 \pm 0.05$} can be deduced.

\emph{Metastable states:}
In the Appendix histograms are shown of the average energy per spin and the modulus squared of the complex XY order parameter for 
increasing system sizes. While the energy histograms are always Gaussian and 
become narrower as the system size increases, the histograms of the order parameter have asymmetric tails 
and are irregularly shaped for large system sizes. This is an indication of metastable states.
% Equilibration but no thermalization as the equilibrated value of the structure factor depends on the initial condition
% of the Monte Carlo run.!!!

\emph{Universality class of the quantum critical point:}
The short-range model has a quantum critical point which is believed to be
in the 3d XY universality class \cite{Blankschtein1984}. 
The XY order parameter is the result of the interplay between frustration on the triangular lattice and quantum fluctuations.
So the effective model at the quantum critical point is that of a \emph{ferromagnetic} XY model that undergoes a clock-order symmetry breaking
transition where the anisotropy does not affect the critical behaviour \cite{Blankschtein1984}. For the long-range ferromagnetic quantum XY model
one expects mean-field critical exponents for $\alpha < \alpha_u = 10/3$ \cite{Dutta2001}. 
While the quality of the data is not good enough to determine the critical exponents directly,
it can be shown that mean-field critical exponents give a better data collapse than for 
example the 3d XY exponents (see Fig. \ref{fig:crit_exp_AFM} in the Appendix),
which is consistent with the above predictions.
From the data collapse the critical field can be determined as $\Gamma_c = 1.15 \pm 0.05$.

\begin{figure}
\includegraphics[width=1.0\linewidth]{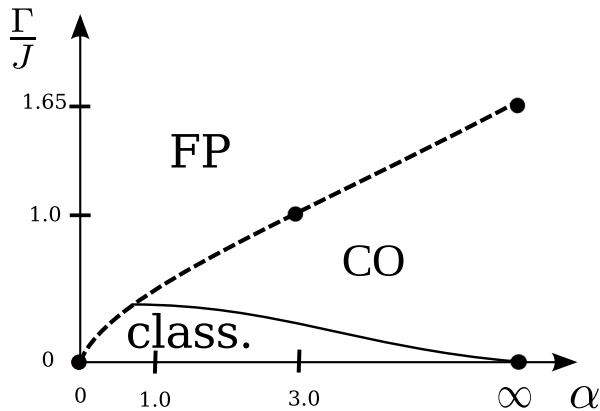}
\caption{Semiquantitative phase diagram for the long-range transverse-field Ising AFM 
on the triangular lattice. {\bf FP:} fully x-polarized phase, {\bf CO:} clock-ordered phase,
{\bf class.:} region dominated by classical ground states.}
\label{fig:AFM_phase_diagram}
\end{figure}

\begin{figure}
\includegraphics[width=1.0\linewidth]{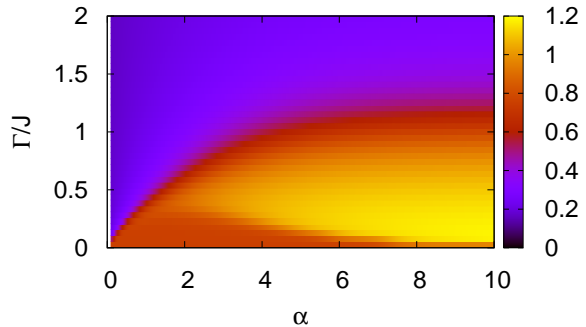}
\caption{Modulus $|m|$ of the complex $XY$ order parameter $m \equiv (m_A + m_B e^{i 2\pi/3} + m_C e^{-i 2\pi/3})/\mathcal{C}$
with $\mathcal{C}=3/2$ computed with Lanczos exact diagonalization for a hexagon of 19 spins.}
\label{fig:alphascan}
\end{figure}

\emph{Semiquantitative phase diagram:}
The intuition from mean field theory and the results of this section can be combined into a
semiquantitative phase diagram (Fig. \ref{fig:AFM_phase_diagram}). 
There we denote by dots the known results for the nearest neighbour model ($\alpha \rightarrow \infty$), the results
of the present simulation ($\alpha = 3$), and the infinite-range model ($\alpha=0$).
The clock symmetry broken phase that extends
for the short-range model between $\Gamma=0^+$ and $\Gamma_c(\alpha=\infty)=1.65 \pm 0.5$ \cite{Isakov2003} persists for 
long-range interactions, with a critical field that decreases with decreasing $\alpha$ as the additional 
frustration due to long-range AFM interactions destabilizes order. For $\alpha=3$ the critical point is located
at $\Gamma_c(\alpha=3) = 1.05 \pm 0.05$, most likely with mean-field critical exponents. For $\alpha=2$ we find that the critical
point is around $\Gamma_c(\alpha=2) \approx 0.8$. 
Whether the ordered phase for $\alpha < 2$ is still clock-ordered is not clear. 
At any rate, for $\alpha \le \alpha_c$ with $\alpha_c \approx 1$ the Fourier transformation of the interactions
becomes flat so that there is no longer a preferred state dictated by the lattice structure that quantum fluctuations can select.
The infinitely coordinated classical Ising AFM at $(\alpha=0, \Gamma=0)$ has an exponential ground state degeneracy:
All states with $S_{\text{tot}}^z = \sum_i S_i^z = 0$ are ground states. Upon introducing a transverse field, the model at $\alpha=0$
turns into an anisotropic variant of the Lipkin-Meshkov-Glick model  \cite{Ribeiro2008} (see Eq. \eqref{eq:LMG}).
From exact diagonalization 
for 19 spins (see Fig. \ref{fig:alphascan}) and from the behaviour of the structure factor at small fields 
in Fig. \ref{fig:sqzz_AFM} one can conclude that for long-range interactions there is a threshold for quantum fluctuations to establish order.
The continuous line in \mbox{Fig. \ref{fig:AFM_phase_diagram}} corresponds to this expectation of a classically dominated region,
where the possible phases of the classical model ($\Gamma=0$) extend to finite values of the transverse field $\Gamma$.
However, with the caveats pointed out in the main text, it should be stressed 
that this phase diagram remains a semiquantitative one.

\section{Conclusion}

The ground state phase diagrams of ferromagnetic and 
antiferromangetic long-range transverse field Ising models on the triangular lattice have been examined. 
The critical transverse field strength $\Gamma_c$ at which the transition from the fully x-polarized phase to the clock-ordered phase
occurs was determined by two independent procedures: by data collapse and by extrapolation of the crossing points of the Binder cumulants.
The results of both methods agree within the error bars.
For the ferromagnet with dipolar decay exponent, $\alpha=3$, the critical field 
is located at 
$\Gamma_c = (9.6 \pm 0.1)J$ (from data collapse) or
$\Gamma_c = (10.0 \pm 0.4)J$ (from extrapolation), which is only slightly below the value from mean field theory.
This is remarkable as the critical field is not a universal quantity. 
%From this it can be inferred that the mean-field phase boundary gradually becomes the exact
%phase boundary as $\alpha$ decreases from 3 to 2.
%However, for finite-size systems, the approach of the critical field to its value in the thermodynamic 
%limit is very slow for $\alpha < 3$.
For the antiferromagnet, there is strong indication for a quantum phase transition from the fully x-polarized phase to a clock ordered phase
at 
$\Gamma_c = (1.15 \pm 0.05)J$ (from data collapse) or
$\Gamma_c = (1.05 \pm 0.05)J$ (from extrapolation).

It remains an open problem to investigate the transition to the clock-ordered 
phase from the side of small $\Gamma$. 
The simulation results indicate that the classical ground states at zero field extend to finite
field, before they finally yield to the strength of quantum fluctuations, which results in the 
clock-ordered phase. The QMC algorithm suffers from loss of ergodicity in this region of small transverse field,
which hampered a quantitative study. It might be possible to overcome this problem by 
parallel tempering in $\Gamma$ which lets simulations at small $\Gamma$, which tend to get stuck in some 
region of phase space, profit from the increased ability to explore phase 
space that simulations at larger $\Gamma$ possess.
Another open issue is the finite-temperature phase diagram of the long-range AFM, 
in particular whether there is a Kosterlitz-Thouless phase 
transition as in the short-range AFM.

The results presented in this paper may be helpful for future ion trap experiments.

\section*{ACKNOWLEDGMENTS}
The author thanks A. Muramatsu for helpful discussions and support for the completion of this work.
Important suggestions by T. Roscilde are also acknowledged.

\newpage

\section{Appendix}

\subsection{Simulation cell}
\begin{figure}[hb!]
 \includegraphics[width=0.8\linewidth]{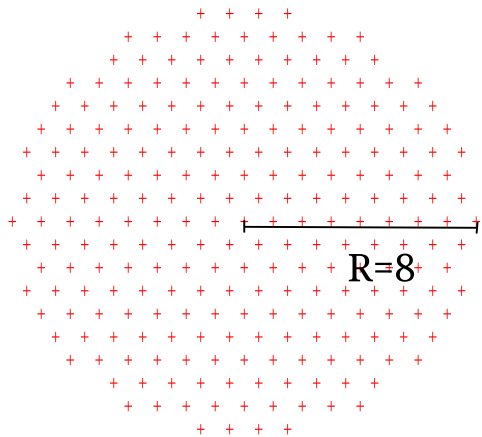}
 \caption{Simulation cell with open boundary conditions with radius $R=8$.}
 \label{fig:simcell}
\end{figure}

For radius $R\le 6$ the shape of the simulation cell is that of a hexagon centered
at a lattice site; for \mbox{$R \ge 7$} it has a circular shape similar to the experimental 
realization in a Penning trap (see Fig. \ref{fig:simcell}).

\subsection{Histograms for energy and XY order parameter}
\begin{figure}
 \begin{tabular}{cc}
      \subfloat[R=3, $|m|^2$]{\includegraphics[width=0.2\textwidth]{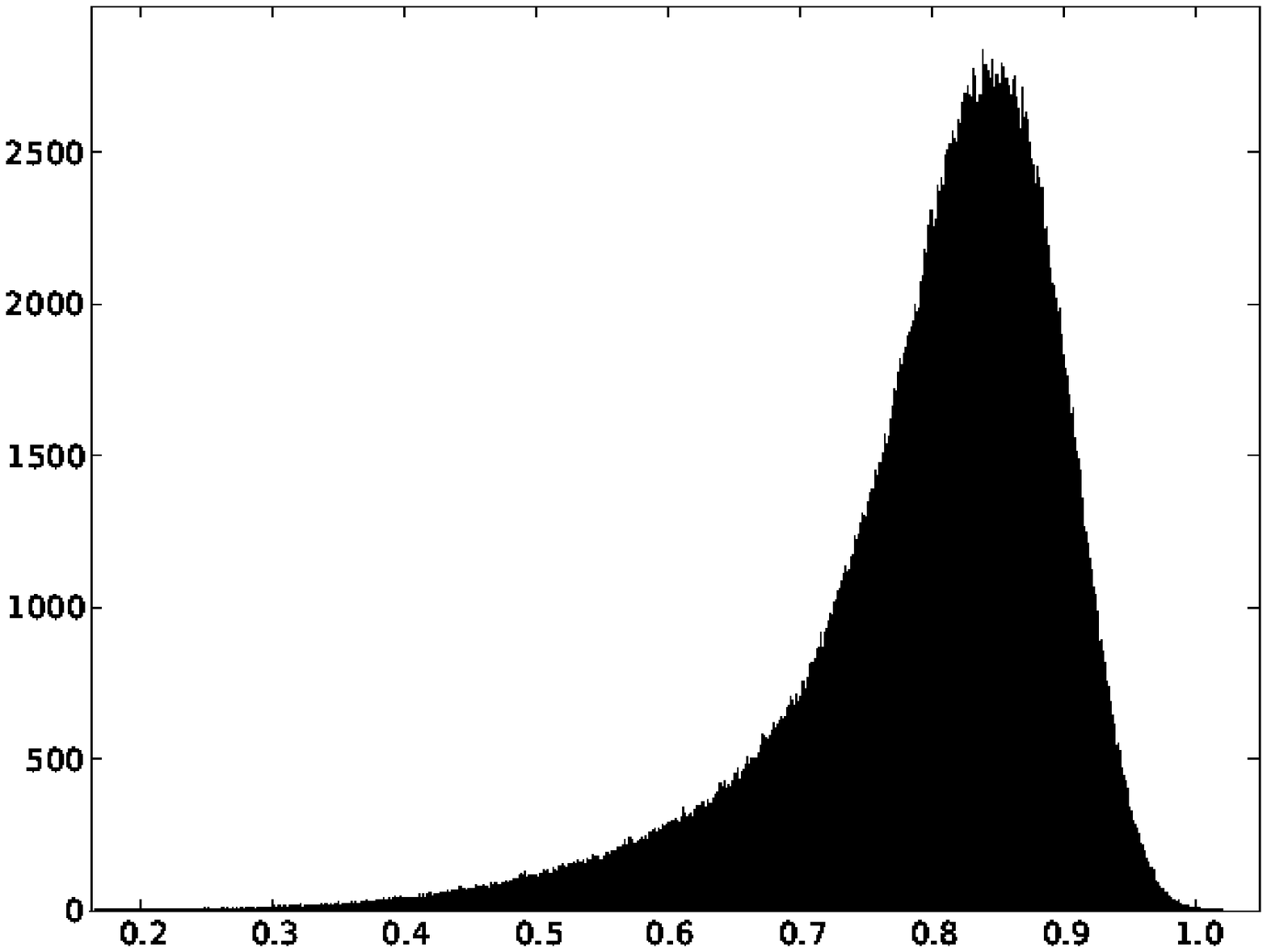}}
    & \subfloat[R=3, energy]{\includegraphics[width=0.2\textwidth]{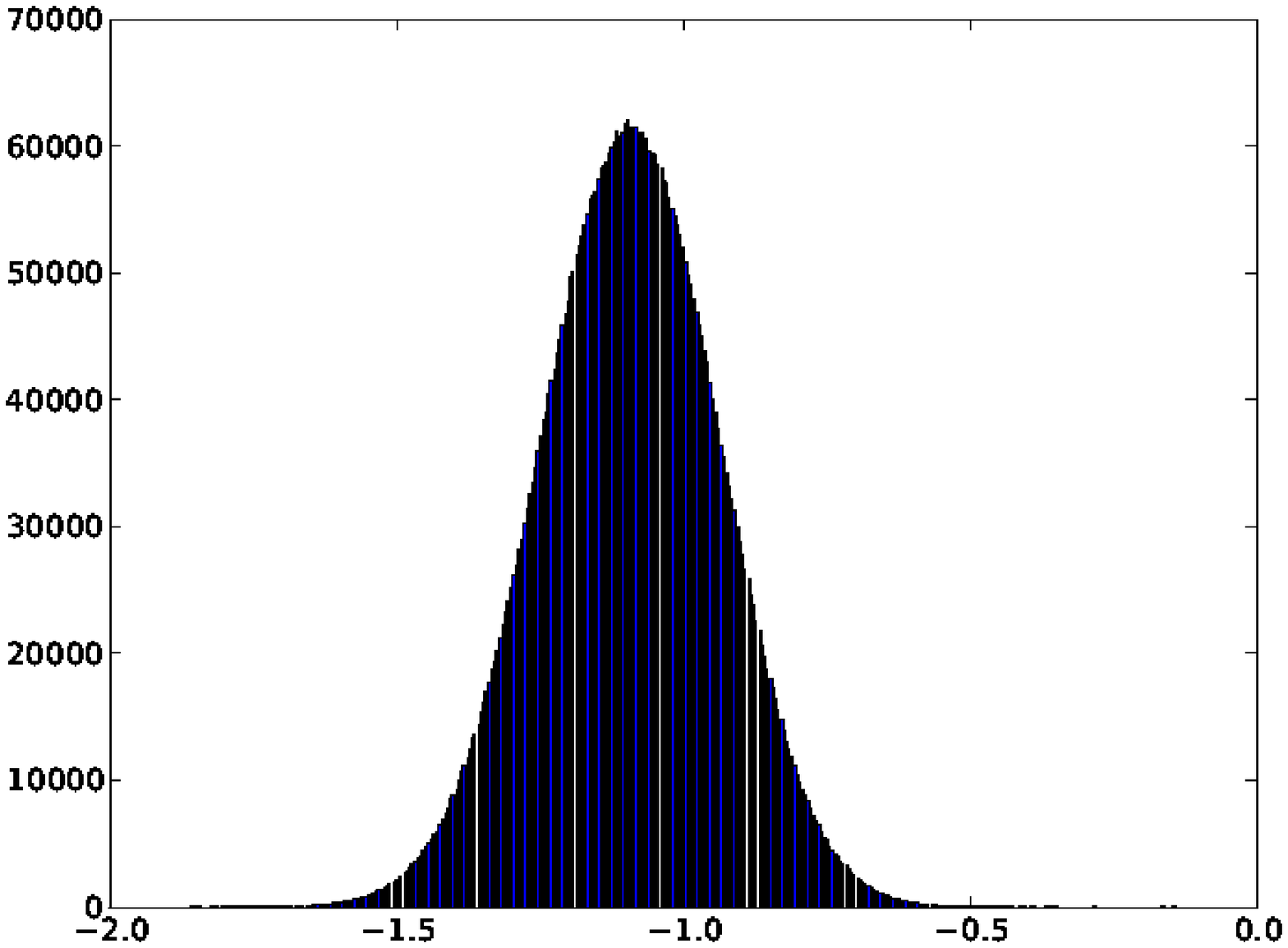}} \\
      \subfloat[R=4,  $|m|^2$]{\includegraphics[width=0.2\textwidth]{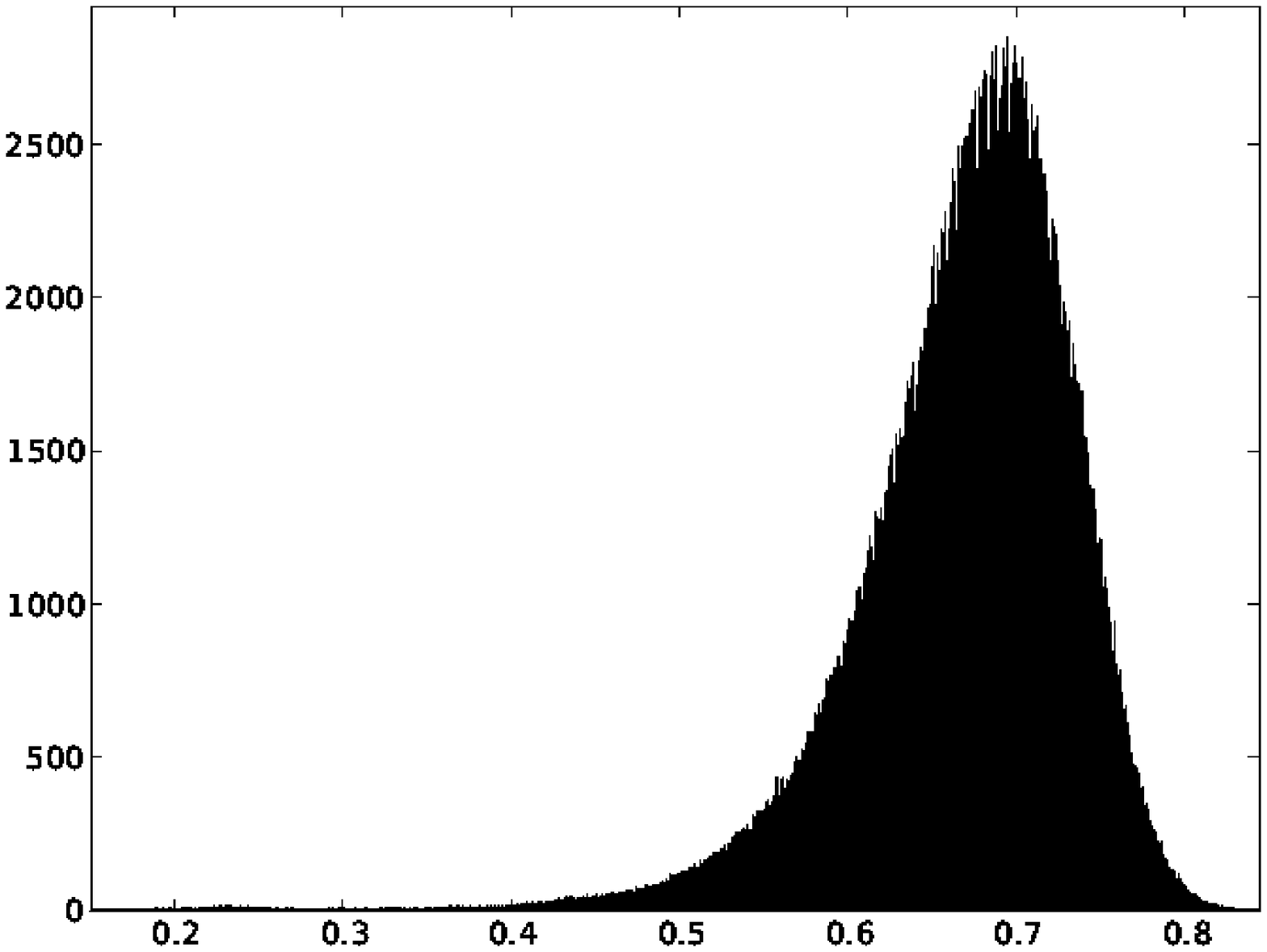}}
    & \subfloat[R=4, energy]{\includegraphics[width=0.2\textwidth]{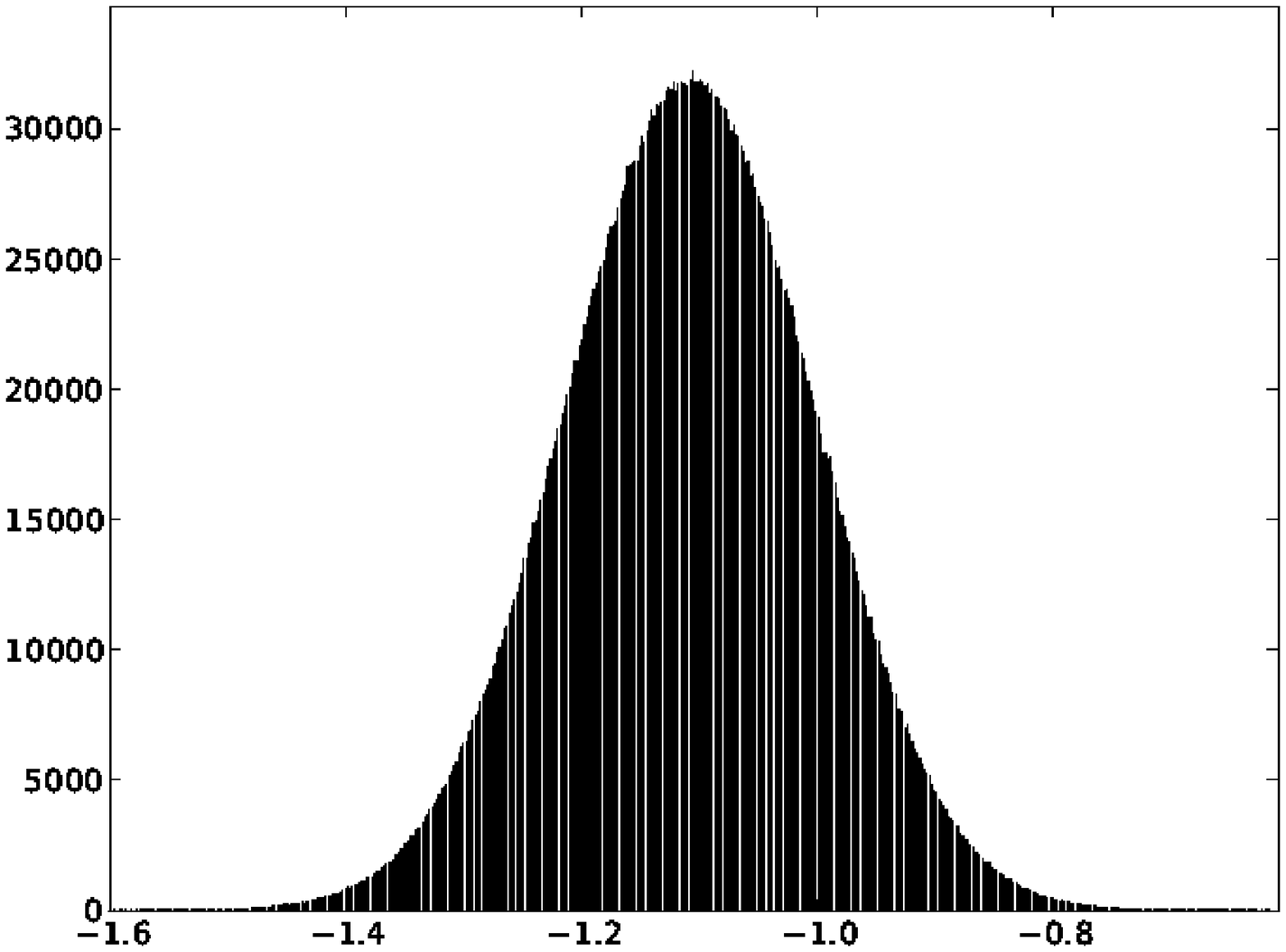}} \\
      \subfloat[R=5,  $|m|^2$]{\includegraphics[width=0.2\textwidth]{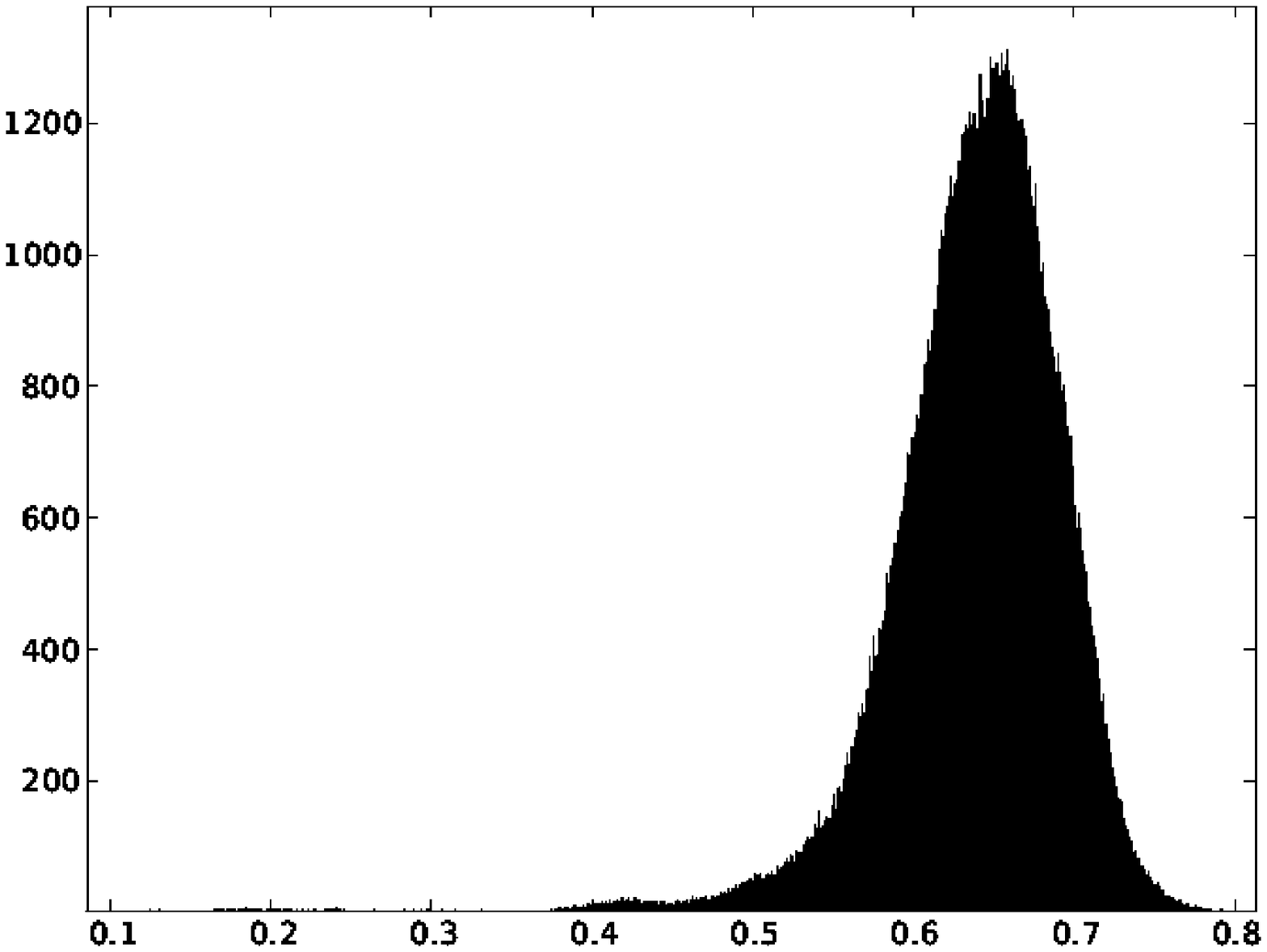}}
    & \subfloat[R=5, energy]{\includegraphics[width=0.2\textwidth]{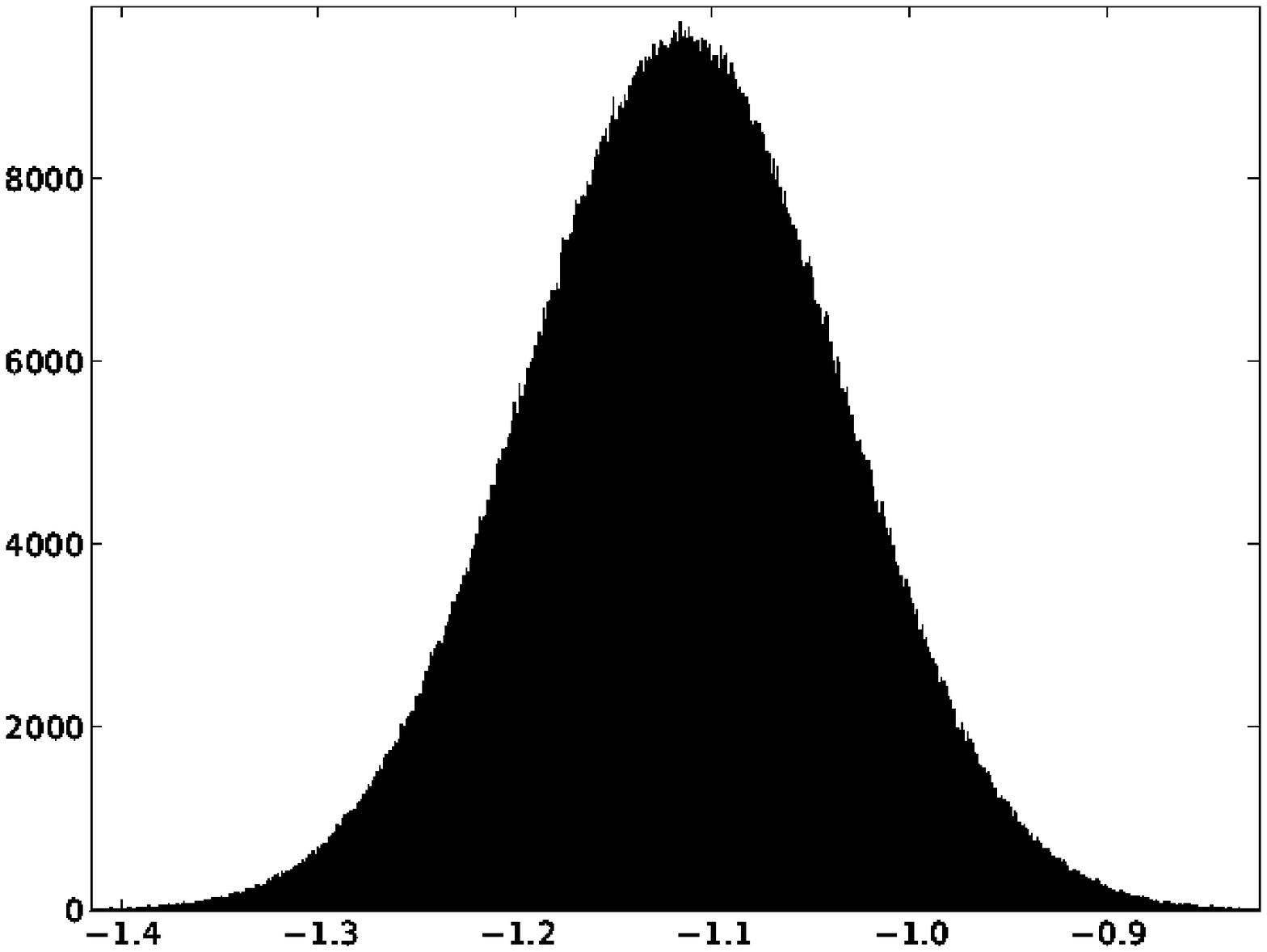}} \\
      \subfloat[R=6, $|m|^2$]{\includegraphics[width=0.2\textwidth]{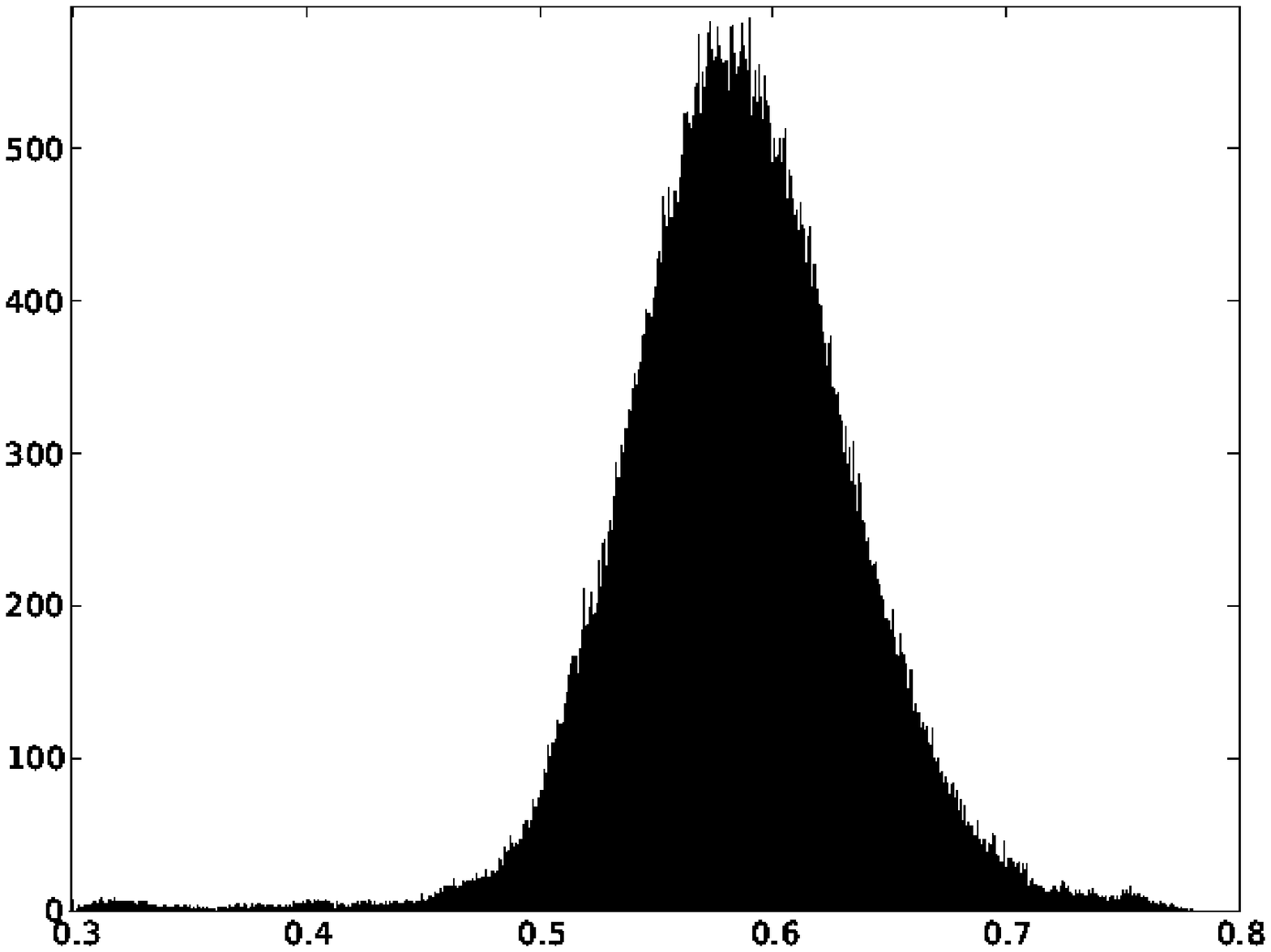}}
    & \subfloat[R=6, energy]{\includegraphics[width=0.2\textwidth]{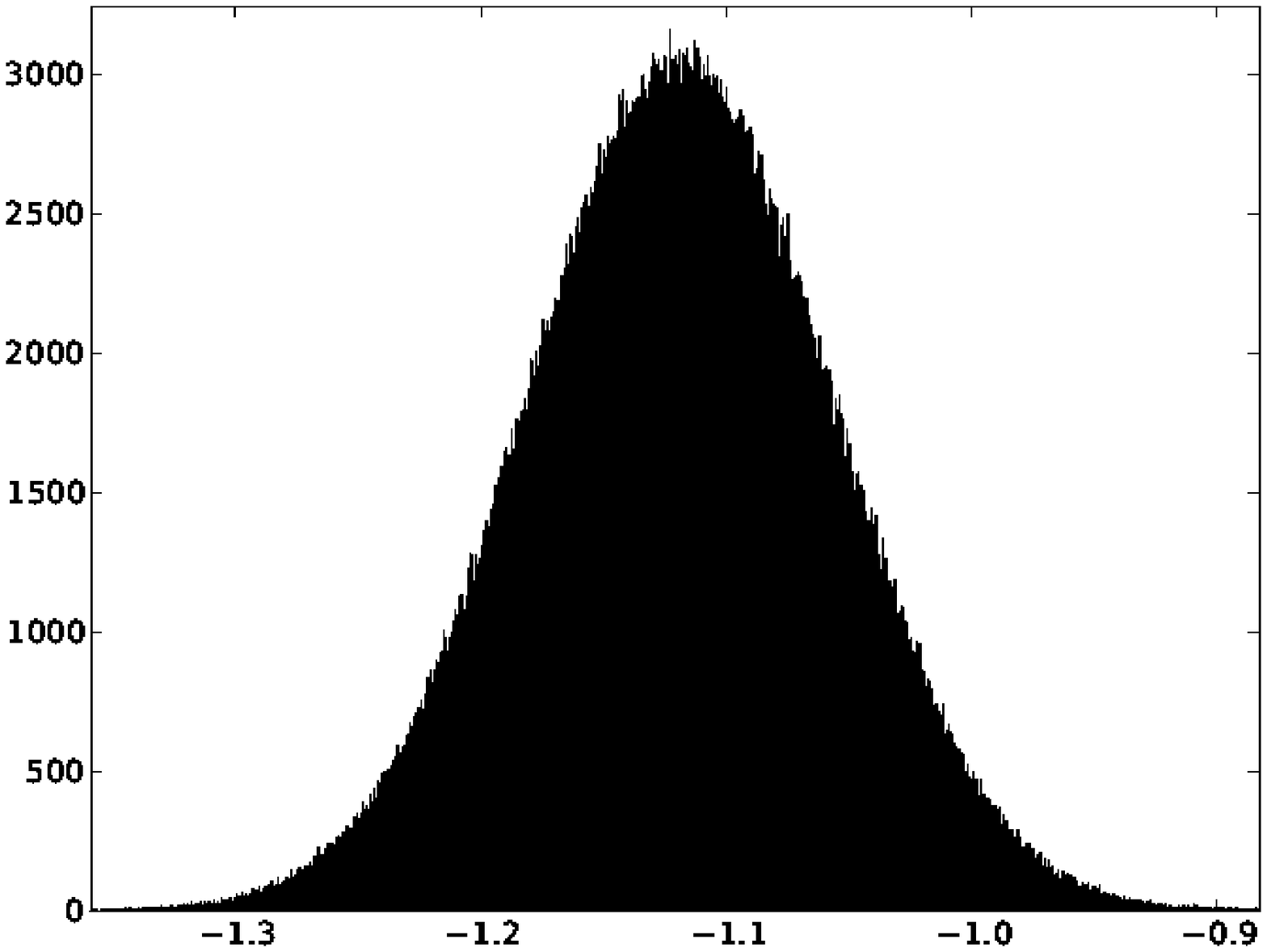}} \\
      \subfloat[R=7,  $|m|^2$]{\includegraphics[width=0.2\textwidth]{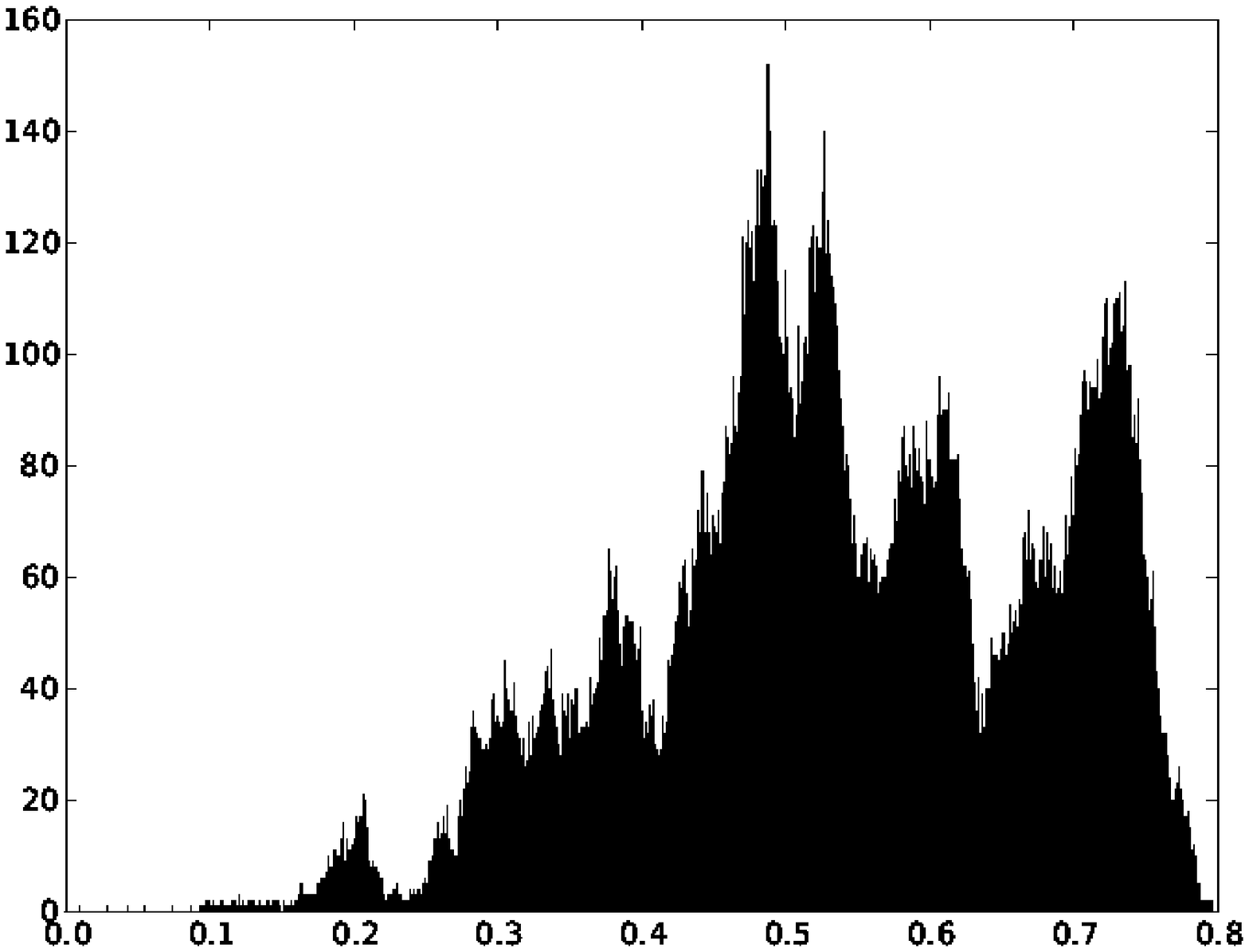}}
    & \subfloat[R=7, energy]{\includegraphics[width=0.2\textwidth]{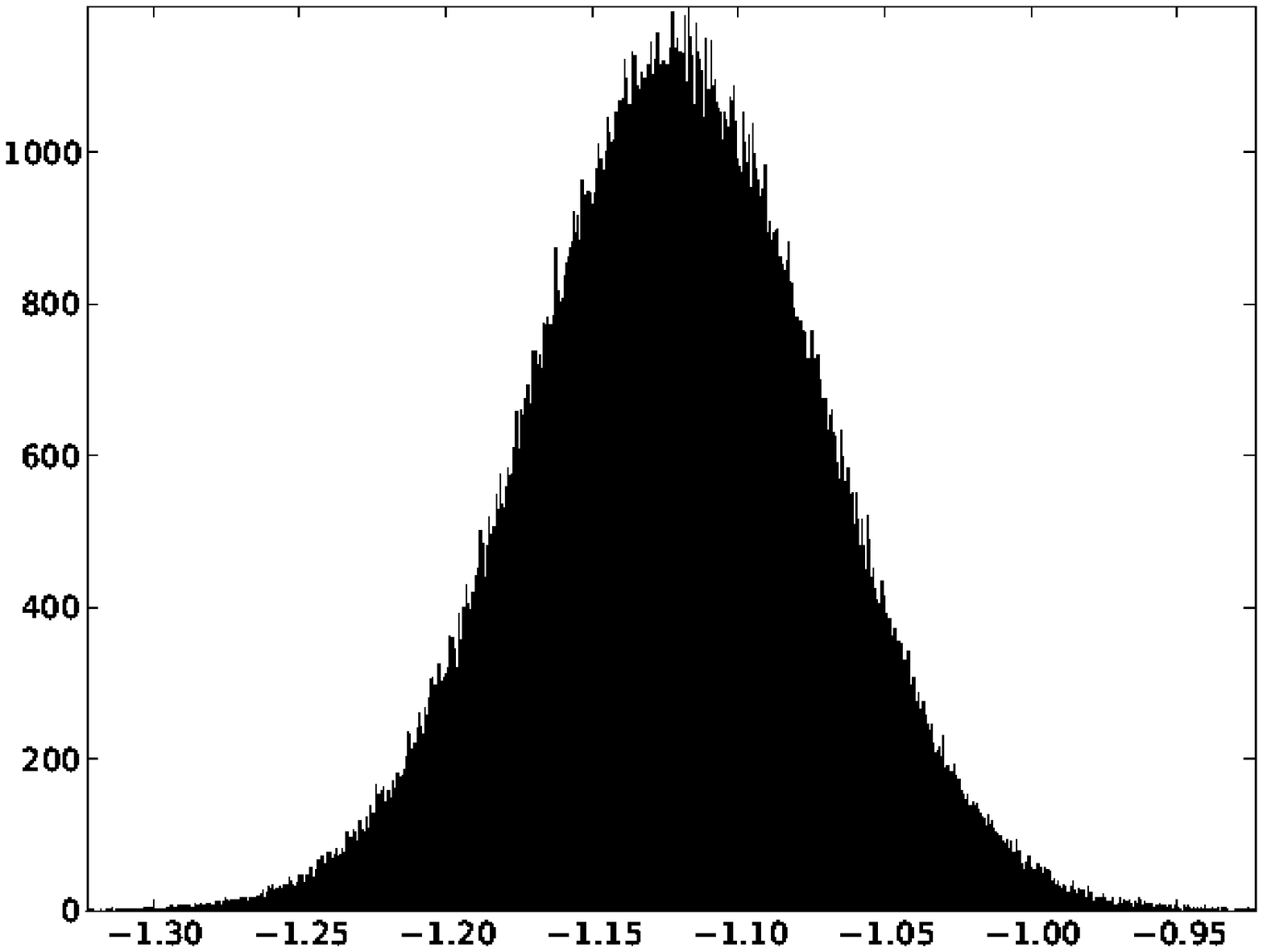}} \\
%   \subfloat[R=8,  $|m|^2$]{\includegraphics[width=0.2\textwidth]{../figures/histLR3R8AFM_orderparam}}
%     & \subfloat[R=8, energy]{\includegraphics[width=0.2\textwidth]{../figures/histLR3R8AFM_energy}} \\
\end{tabular}
 \caption{Histograms for the average energy per spin and the modulus squared $|m|^2$ of the complex XY order parameter for $(\alpha=3, \Gamma = 0.7)$
 and different radii $R$ of the simulation cell.}
 \label{fig:histograms}
\end{figure}

As pointed out in the main text, the error bars in the clock ordered phase are not reliable. 
The reason is that the histograms of the modulus squared $|m|^2$ of the complex XY order parameter
are not Gaussian, but exhibit long tails, or even show no well-defined shape at all due to metastabilities.
This is illustrated by the histograms in Fig. \ref{fig:histograms} where the distribution 
of $|m|^2$ and the distribution of the energy per spin are contrasted. 
The bin size for the histograms is chosen such that they appear continuous, 
and the units of the $y-$axis are arbitrary.
The histograms in Figs. \ref{fig:histograms}(a),(b), and (c) are examples 
of distributions with asymmetric tails. Fig. \ref{fig:histograms}(i) shows a distribution in the presence
of strong metastabilities which can no longer be described by a mean value and a standard deviation.
On the other hand, the histograms of the energy per spin are always Gaussian
and their width shrinks with increasing system size, as is to be expected according to the Central Limit Theorem.

\subsection{Data collapse for the AFM quantum critical point}
\begin{figure}[hb!]
 \includegraphics[width=1.0\linewidth]{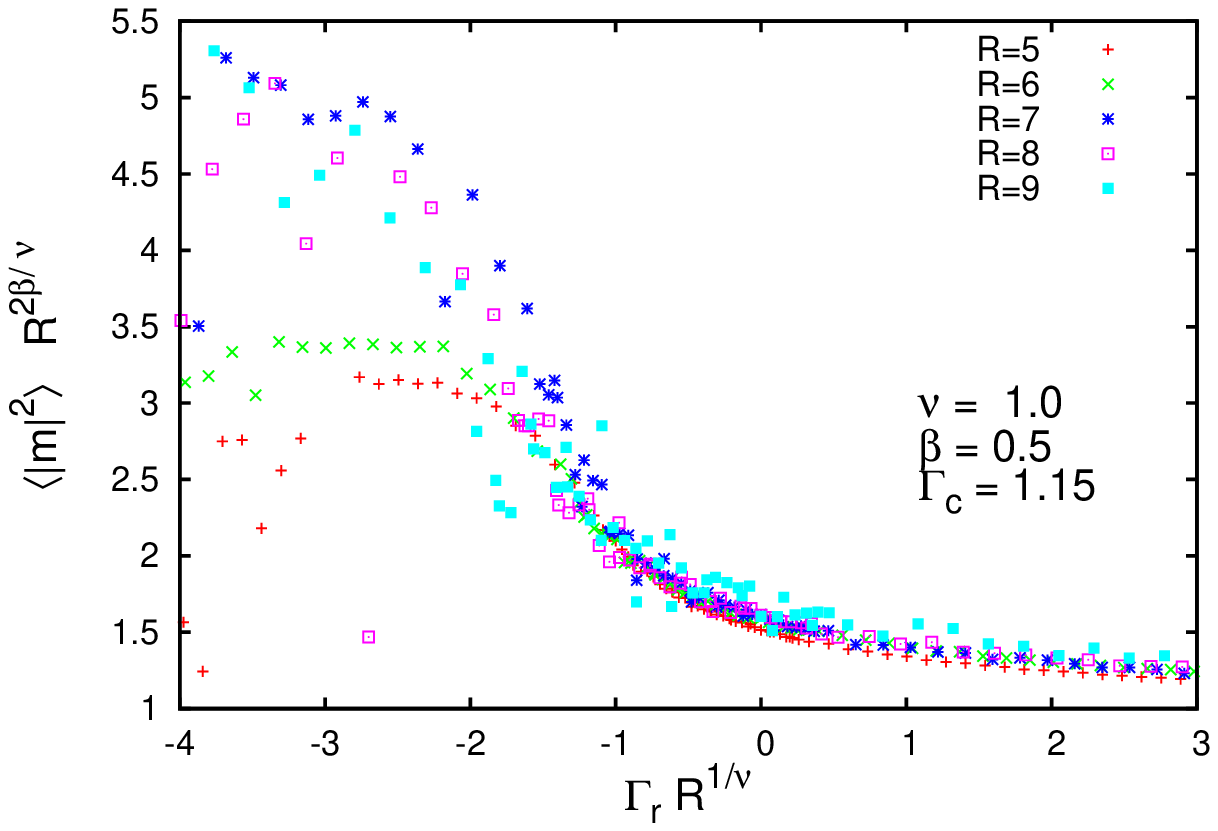}  \\%AFM_alpha3_FSS_Gammac115
 \includegraphics[width=1.0\linewidth]{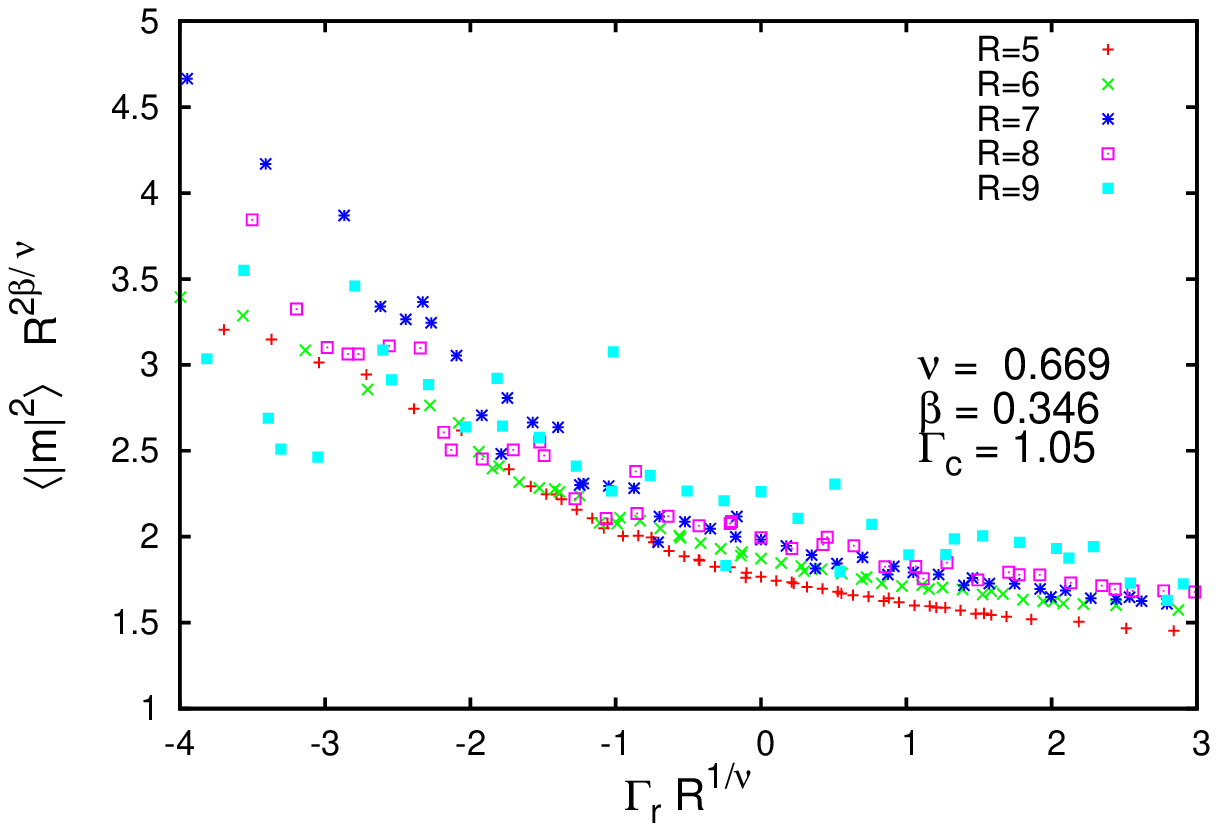}
 \caption{Data collapse for $\alpha=3$ with mean-field exponents and $\Gamma_c = 1.15$ (upper panel) and
 data collapse with 3d XY exponents and $\Gamma_c = 1.05$ (lower panel).
 $\Gamma_r \equiv (\Gamma - \Gamma_c)/ \Gamma_c$ is the reduced field.}
 \label{fig:crit_exp_AFM}
\end{figure}

It has been detailed in the main text that the short-range transverse field Ising 
AFM on the triangular lattice is believed to have a quantum critical point in the 3d XY universality class,
whereas the quantum critical point for the long-range counterpart should be in the mean field 
universality class. A comparison of the data collapse with one or the other set of exponents 
for the long-range AFM with $\alpha=3$ is shown in Fig. \ref{fig:crit_exp_AFM}.
With mean-field exponents, $\nu=1$ and $\beta=\frac{1}{2}$, the best achievable data collapse 
is obtained for \mbox{$\Gamma_c = 1.15$}. 
The best achievable data collapse with 3d XY exponents, $\nu=0.669$ and $\beta=0.346$, 
is shown in the lower panel. Slightly different transverse fields lead to a similar unsatisfactory 
data collapse in that case.
 
\end{document}